\documentclass[12pt,a4paper]{article}
\usepackage[dvips]{graphicx,color}
\usepackage{times}
\usepackage{xcolor}
\usepackage{cite}
\usepackage{float}
\usepackage[%
  colorlinks=true,
  urlcolor=blue,
  linkcolor=green,
  citecolor=blue]{hyperref}
\usepackage[utf8]{inputenc}
\usepackage{amsmath}
\usepackage{amsfonts}
\usepackage{makecell}
\usepackage{amssymb}
\usepackage{graphicx}
\usepackage{pgfplots}
\usepackage{subcaption}
\usepackage{caption}
\pgfplotsset{compat=1.18} 
\usepackage{multirow}
\usepackage{booktabs}
\usepackage[left=2.0cm,right=2.0cm,top=2.5cm,bottom=2.5cm]{geometry}
\begin{document}
\newcommand{\sheptitle}
{Renormalization group evolution induced breaking of $\mu-\tau$ reflection symmetry in MSSM with effects of variation of $tan\beta$}
\newcommand{\shepauthor}
{Chandan Kumar Borah \footnote{ E-mail: cborah528@gmail.com}\ and\ 
 Chandan Duarah \footnote{ E-mail: chandanduarah@dibru.ac.in}}
\newcommand{\shepaddress}
   {Department of Physics, Dibrugarh University,
               Dibrugarh - 786004, India }
\newcommand{\shepabstract}
{\noindent We study the renormalization group (RG) evolution induced breaking of $\mu$--$\tau$ reflection symmetry in the Minimal Supersymmetric Standard Model (MSSM), with a special focus on the effects of varying $\tan\beta \equiv v_u/v_d$, the ratio of MSSM Higgs vacuum expectation values. Starting from an exact $\mu$--$\tau$ reflection symmetry imposed at a high flavor symmetry scale $\Lambda_{\text{FS}}$, we run the complete set of coupled RGEs for neutrino masses, mixing angles, and CP-violating phases down to the electroweak scale, imparting perturbation to the symmetry. We consider a specific value of the SUSY breaking scale, $\Lambda_s=7\ TeV$ during the run. By choosing suitable free parameters at the high-energy scale, we reproduce the low-energy experimental constraints on neutrino observables consistent with $3\nu$ global analysis data. We then examine how the breaking of $\mu$--$\tau$ reflection symmetry is influenced by different values of $\tan\beta$, considering three benchmark choices. In addition, the analysis is performed for both normal ordering (NO) and inverted ordering (IO) of neutrino masses to highlight potential differences in their RG running behavior.
	\bigskip
	
	\noindent
	{\bfseries Keywords: Lepton mixing, $\mu-\tau$ reflection symmetry,  Renormalization group running, MSSM.} 
	\vspace{1.2cm}
\noindent
\pagebreak}
\begin{titlepage}
\begin{flushright}
\end{flushright}
\begin{center}
{\large{\bf\sheptitle}}
\bigskip\\
\shepauthor
\\
\mbox{}\\
{\it\shepaddress}\\
\vspace{.5in}
{\bf Abstract}
\bigskip
\end{center}
\setcounter{page}{0}
\shepabstract
\end{titlepage}

\section{Introduction}
\indent Neutrino oscillations establish that neutrinos are massive particles and their flavor states are linear combinations of mass eigenstates. This mixing between flavor and mass eigenstates is described by the Pontecorvo–Maki–Nakagawa–Sakata (PMNS) matrix. In general, the PMNS matrix is expressed in terms of three mixing angles—the solar angle $\theta_{12}$, the atmospheric angle $\theta_{23}$, and the reactor angle $\theta_{13}$ along with one Dirac CP-violating phase $\delta$, two Majorana phases $\alpha$ and $\beta$ and three unphysical phases $\phi_1$, $\phi_2$ and $\phi_3$. It is given by \cite{PDG}
\begin{equation}
    U=P_1VP_2 \label{U},
    \end{equation}  with
\begin{equation}
    V=\begin{bmatrix}
  c_{12}c_{13}&s_{12}s_{13}&s_{13}e^{-i\delta}\\-s_{12}c_{23}-c_{12}s_{23}e^{i\delta}&c_{12}c_{23}-s_{12}s_{23}e^{i\delta}&c_{13}s_{23}\\s_{12}s_{23}-c_{12}s_{13}e^{i\delta}&-c_{12}s_{23}-s_{12}c_{23}e^{i\delta}&c_{13}c_{23} \end{bmatrix},
  \label{para}
\end{equation}
 $P_1=Diag(e^{i\phi_1},e^{i\phi_2},e^{i\phi_3})$ and $P_2=Diag(e^{i{\alpha}},e^{i{\beta}},1)$. In the matrix V, $s_{ij}=\sin\theta_{ij}$ and $c_{ij}=\cos\theta_{ij}$ ($ij=12,23,13$).

The values of $\theta_{12}$ and $\theta_{13}$ are determined experimentally with a good precision \cite{SKT2K, SK24, SNO, Daya, Nova1, Nova2, IceCube, T2k1, T2k2}. The recent JUNO measurement reports $\sin^2\theta_{12} = 0.3092 \pm 0.0087$ and $\Delta m_{21}^2 = (7.50 \pm 0.12)\times 10^{-5}\ \text{eV}^2$ for the normal mass-ordering scenario \cite{JUNO}. Furthermore, the experimentally determined values of the elements in the PMNS matrix reveal that the $\mu$ and $\tau$ flavor components are approximately equal. This near equality suggests the existence of an underlying symmetry in the lepton sector referred to as $\mu$–$\tau$ flavor symmetry in neutrino mixing. A general prediction of $\mu$–$\tau$ flavour symmetry is that the atmospheric mixing angle $\theta_{23}$ is maximal. In the more specific case of $\mu$–$\tau$ reflection symmetry, the prediction extends to the CP-violating phase, yielding $\delta = \pi/2$ or $3\pi/2$. Notably, this prediction is consistent with recent results from the T2K and NO$\nu$A experiments \cite{T2k1, T2k2, Nova1, Nova2}. The near-maximal value of $\delta$ is also reflected in the recent global analysis of neutrino oscillation data \cite{GlobAnal} in the inverted order (IO) scenario, thereby bringing reflection symmetry into particular focus in current studies. Hence, it has been the subject of extensive investigation in recent years \cite{RRS1, RRS2, RRS3, RRS12, RRS4, RRS5, RRS6, RRS13, NNath, dirac, RRS9, VVV1, VVV2, RRS10, M.Kashav}. 

The $\mu$–$\tau$ reflection symmetry was first introduced by Harrison and Scott \cite{RS1}. They formulated a specific parametrization of the mixing matrix given by
\begin{equation}
    V_{HS}=\begin{bmatrix}
        u_1 & u_2 &u_3 \\ v_1 &v_2 &v_3 \\ v_1^* & v_2^* & v_3^*
    \end{bmatrix},
    \label{HS}
\end{equation}
where $u_i$ are real and $v_i$ are complex parameters and the matrix is consistent with the equality $|U_{\mu i}| = |U_{\tau i}|$. The symmetry operation consists of interchanging the $\mu$ and $\tau$ flavor components while simultaneously applying charge conjugation:
$$\nu_e \rightarrow \nu_e^c,\  \nu_\mu \rightarrow \nu_\tau^c,\  \nu_\tau \rightarrow \nu_\mu^c.$$
The corresponding invariant neutrino mass matrix is given by 
\begin{equation}
    M=\begin{bmatrix}
        M_{ee} & M_{e\mu} & M_{e\tau}^* \\ M_{e\mu} & M_{\mu\mu} &M_{\mu\tau} \\ M_{e\tau}^* &M_{\mu\tau} & M_{\mu\mu}^*
    \end{bmatrix},
\end{equation}
where the elements $M_{ee}$ and $M_{\mu\tau}$ are real.

As reflection symmetry predicts two maximal values of the Dirac phase, we identify the following two cases:
\begin{itemize}
\item \textbf{Case I:} $\theta_{23} = \pi/4$, $\delta = \pi/2$,
\item \textbf{Case II:} $\theta_{23} = \pi/4$, $\delta = 3\pi/2$.
\end{itemize}
with the following two sets of maximal values for the Majorana phases and one of the unphysical phase in the parametrization of PMNS matrix given in Eq (\ref{U}) are consistent with $\mu$–$\tau$ reflection symmetry.

For Case I,
\begin{equation}
\alpha = \beta = \tfrac{3\pi}{2}, \quad \phi_1 = \tfrac{\pi}{2}, \quad \phi_2 = \phi_3 = 0,
\label{ch1}
\end{equation}

and for Case II,
\begin{equation}
\alpha = \beta = \tfrac{\pi}{2}, \quad \phi_1 = \tfrac{3\pi}{2}, \quad \phi_2 = \phi_3 = 0.
\label{ch2}
\end{equation}

These specific phase choices were first employed in the standard parametrization of the PMNS matrix in Ref.~\cite{CDuarah}. Using these specific maximal choices for the CP phases in the standard parametrization of the mixing matrix given in Eq. (\ref{U}), we can reproduce the reflection symmetric matrix as given by Harrison and Scott in Eq.(\ref{HS}). 

The global analysis of the neutrino oscillation data indicates that the mixing parameters deviate from the maximal values (Table \ref{GA}). This makes it essential to study perturbations of $\mu$–$\tau$ reflection symmetry using different theoretical approaches. In our previous work \cite{CBorah}, we investigated the breaking of $\mu$–$\tau$ reflection symmetry due to the Renormalization group (RG) running effects. Assuming the symmetry to be exact at the seesaw scale, we adopted the symmetry-predicted maximal values for the relevant parameters in that regime. By solving the renormalization group equations (RGEs) simultaneously, we demonstrated how deviations from maximality arise as the energy is lowered. In that analysis, three mass eigenvalues and two mixing angles served as free parameters at the high scale and by carefully choosing their values, we obtained results consistent with low-energy constraints.
 Similar works about the impact of RG running on $\mu$–$\tau$ reflection symmetry have been analyzed in \cite{RRS1, RRS6, dirac, JZhu, YLZhou, RGERS2}. On the other hand, general RG effects on neutrino parameters are found extensively in \cite{KS, RGE2, 1, R1, R3, R2, poko, babu, R4, JMei, R5, ohl, Dzhang, NNSingh, SGupta, RGERS2, YLZhou, JZhu}
 
\begin{table}[t]
\begin{center}
\begin{tabular}{c cc cc }
\hline
 Without SK atmospheric data & & \\ \hline
\multirow{2}{*}{Parameter}& 
\multicolumn{2}{c}{Normal Ordering}&
\multicolumn{2}{c}{Inverted Ordering} \\
\cline{2-5}
 &Best-fit Value  & $3\sigma$ & best-fit value & $3\sigma$ \\ \hline
 $\theta_{12}$         & 33.68    &  31.63-35.95   & 33.68    &  31.63-35.95         \\
 $\theta_{23}$         & 48.5    & 41.0-50.5       & 48.6    & 41.4-50.6      \\
 $\theta_{13}$         & 8.52    & 8.18-8.87        & 8.58    & 8.24-8.91    \\ 
 $\delta$         & 177     & 96-422  & 285    & 201-348         \\
 $\Delta m^2_{21}(/10^{-5}eV^2)$  & 7.49   & 6.92-8.05 & 7.49   & 6.92-8.05        \\
 $\Delta m^2_{32}(/10^{-3}eV^2)$  & 2.534 & 2.463-2.606 & -2.510     & -2.584-2.438          \\ \hline
  With SK atmospheric data & & \\ \hline
 $\theta_{12}$         & 33.68    &  31.63-35.95   & 33.68    &  31.63-35.95         \\
 $\theta_{23}$         & 43.3    & 41.3-49.9       & 47.9    & 41.5-49.8      \\
 $\theta_{13}$         & 8.56    & 8.19-8.89        & 8.59    & 8.25-8.93    \\ 
 $\delta$         & 212     & 124-364  & 274    & 201-335         \\
 $\Delta m^2_{21}(/10^{-5}eV^2)$  & 7.49   & 6.92-8.05 & 7.49   & 6.92-8.05        \\
 $\Delta m^2_{32}(/10^{-3}eV^2)$  & 2.513 & 2.451-2.578 & -2.484     & -2.547-2.421          \\ \hline
\end{tabular}
\end{center}
\caption{The best-fit values and 3$\sigma$ allowed ranges of neutrino oscillation parameters in NO and IO obtained from global analysis \cite{GlobAnal}.}
\label{GA}
\end{table}

In this work, we investigate the role of $\tan\beta$ on RG-induced breaking of $\mu$–$\tau$ reflection symmetry in the minimal supersymmetric standard model (MSSM). Starting from exact $\mu$–$\tau$ reflection symmetry predicted maximal values at $\Lambda_{FS}$, we run the full set of coupled RGEs for neutrino parameters, with proper matching across the SUSY breaking scale $\Lambda_{s}$. We consider $\Lambda_s = 7~\text{TeV}$ and evolve the parameters down to the electroweak scale $\Lambda_{EW}$. Note that from $\Lambda_{FS}$ to $\Lambda_s$, we work within the minimal supersymmetric standard model (MSSM). Below the scale $\Lambda_s$, the effective theory is the Standard Model, where we additionally take into account the RGE for the Higgs quartic coupling. By selecting appropriate high-scale inputs for the free parameters, we aim to reproduce the experimentally allowed $3\sigma$ ranges of neutrino oscillation parameters at low energies.

We adopt the RGEs for the neutrino mixing parameters derived in our previous work \cite{CBorah}. In addition, we also consider the RGEs for the gauge and Yukawa couplings. These constitute a set of nineteen coupled differential equations, which we solve numerically using {\tt Python} code. For the high-energy input values of the gauge and Yukawa couplings, we employ a bottom-up approach, i.e, starting from their values at the electroweak scale, we evolve them upward by solving their RGEs, following the procedure outlined in our earlier work. However, in the present analysis, the high-energy value obtained accordingly differs depending on the choice of $\tan\beta$. 

We then analyze the effect of varying $\tan\beta$ on symmetry breaking by considering its three benchmark values ($10,\ 30,$ and $58$). Our results demonstrate that larger $\tan\beta$ values enhance the deviations of $\theta_{23}$ and $\delta$ from their symmetric predictions, with non-negligible effects also observed in $\theta_{12}$ and $\theta_{13}$. Furthermore, we study both normal ordering (NO) and inverted ordering (IO) of neutrino masses and we show how the pattern of RG-induced corrections depends on the choice of mass ordering.

Although the study of symmetry breaking induced by renormalization group (RG) effects has been extensively discussed in the literature, our work offers a distinctive perspective in this context. We carried out detailed analytical and numerical calculations by solving nineteen coupled differential equations to examine the running behaviour of neutrino parameters and the associated symmetry breaking. At the seesaw scale, the theoretical framework is based on the Minimal Supersymmetric Standard Model (MSSM); however, below the SUSY breaking scale, the Standard Model (SM) is adopted. In our previous analysis, we investigated the running effects for three different SUSY breaking scales while keeping the value of $\tan\beta$ fixed. In the present study, we aim to explore the impact of varying $\tan\beta$ for a chosen SUSY breaking scale

The remaining part of this paper is organized as follows. In Section 2, we discuss the one-loop RGEs of neutrino mass eigenvalues and lepton mixing parameters and show the dependence of neutrino parameters on $tan\beta$. In Section 3, we present the numerical analysis and results. Finally, the conclusions are given in Section 5.

\section{Renormalization group evolution of neutrino parameters and the role of $\tan\beta$}
The massive nature of neutrinos necessarily implies the existence of new physics beyond the Standard Model (SM). Various mechanisms have been proposed to generate neutrino masses through suitable extensions of the SM, and the most natural approach among them is through an effective dimension-five Weinberg operator:
\begin{equation}
\mathcal{L}_{\kappa}^{\rm SM} \;=\; \frac{1}{4}\,\kappa_{\alpha\beta}\,(\overline{L^c_\alpha} \, H)\, (L_\beta H) \;+\; \text{h.c.}\,,
\label{eq:Weinberg-operator}
\end{equation}
where \(L\) is the lepton doublet and \(H\) the SM Higgs doublet. After electroweak symmetry breaking (EWSB), one obtains the Majorana mass matrix
\begin{equation}
M_\nu^{\rm SM} \;=\; \frac{v^2}{2}\,\kappa \,,
\qquad
v \equiv \langle H\rangle \simeq 174~\text{GeV}\,.
\label{eq:Mnu-SM}
\end{equation}
In the MSSM, the superpotential contains
\begin{equation}
W \supset \frac{1}{2}\,\kappa_{\alpha\beta}\,(L_\alpha H_u)\,(L_\beta H_u)\,,
\end{equation}
leading to
\begin{equation}
M_\nu^{\rm MSSM} \;=\; v_u^2\,\kappa \,,
\end{equation}

Considering the charged-lepton Yukawa matrix \(Y_l\) is diagonal, \(Y_l=\mathrm{diag}(y_e,y_\mu,y_\tau)\) and denoting \(t\equiv \ln\mu\), the one-loop RGEs of $\kappa$ is given by:

\begin{equation}
 16\pi^2 \frac{d\kappa}{dt}=C\left[\kappa({Y_l}^\dagger Y_l)+({Y_l}^\dagger Y_l)^T\kappa \right]+\alpha \kappa,
 \label{12}
\end{equation}
where
\begin{equation}
    C_{SM}=-\frac{3}{2}, \ \ \ \ \alpha_{SM}\approx2y_\tau^2+6y_t^2-3g_2^2+\lambda
    \label{constSM}
\end{equation}
for SM and
\begin{equation}
    C_{MSSM}=1,\  \  \  \  \alpha_{MSSM}\approx6y_t^2-\frac{6}{5}g_1^2-6g_2^2
    \label{consMSSM}
\end{equation}
for MSSM. In the above expressions, $y_{\tau}$ and $y_t$ are the $\tau$ lepton and top quark Yukawa coupling constants, respectively; $g_1$ and $g_2$ are the gauge coupling constants and $\lambda$ is the Higgs self-coupling constant. 

Using the RGE of the effective neutrino mass operator $\kappa$, the RGEs for the individual neutrino parameters can be obtained by following the standard procedure described in \cite{KS, RGE2, R1, R3}. In our previous work \cite{CBorah}, we derived the complete set of these equations, which are adopted in the present analysis. It is worth noting that the RGEs for the mass eigenvalues in the SM and MSSM exhibit distinct forms, while those for the CP-violating phases differ primarily through the coefficient $C$ appearing in the RGE of $\kappa$. Since the explicit expressions have already been presented in our earlier work, we do not reproduce them here in order to save space. Nevertheless, all of these equations are employed in our numerical analysis. From the RGEs presented there, it is evident that their forms differ between the SM and the MSSM due to the distinct particle content and coupling structure in the two frameworks. Since our objective is to investigate the radiative breaking of the $\mu$–$\tau$ reflection symmetry, assuming it to be exact at a high-energy scale and studying its deviation as the energy scale decreases, it becomes necessary to incorporate both the MSSM and SM evolutions in our analysis. Specifically, the renormalization group running from the high-energy seesaw scale down to the supersymmetry (SUSY) breaking scale is governed by the MSSM RGEs, as supersymmetric particles remain active in this region. Below the SUSY breaking scale, the supersymmetric degrees of freedom decouple, and the evolution is subsequently described by the SM RGEs down to the electroweak scale.

In the SM, the Higgs sector contains a single Higgs doublet with vacuum expectation value (VEV) $v\simeq 174~\text{GeV}$. In terms of v, the Yukawa couplings of the third-generation fermions are expressed as 
\begin{equation}
y_t = \frac{m_t}{v}, \qquad  
y_b = \frac{m_b}{v}, \qquad  
y_\tau = \frac{m_\tau}{v}.
\end{equation}
Thus, in the SM framework, these Yukawa couplings are determined directly by the fermion masses and the single VEV $v$.  

In contrast, the Minimal Supersymmetric Standard Model (MSSM) requires two Higgs doublets, $H_u$ and $H_d$, to give masses separately to up-type and down-type fermions. Their vacuum expectation values are denoted as $v_u$ and $v_d$, respectively, with 
\begin{equation}
\tan\beta \equiv \frac{v_u}{v_d}, \qquad v = \sqrt{v_u^2 + v_d^2} \simeq 174~\text{GeV}.
\end{equation}
The third-generation Yukawa couplings are then modified as 
\begin{equation}
y_t = \frac{m_t}{v \sin\beta}, \qquad  
y_b = \frac{m_b}{v \cos\beta}, \qquad  
y_\tau = \frac{m_\tau}{v \cos\beta}.
\end{equation}

It is evident from these relations that $\tan\beta$ plays a crucial role in determining the strength of Yukawa couplings. For small values of $\tan\beta$, the tau and bottom Yukawa couplings remain relatively suppressed, while for large values of $\tan\beta$, both $y_b$ and $y_\tau$ are significantly enhanced due to the $1/\cos\beta$ factor. On the other hand, the top Yukawa coupling is proportional to $1/\sin\beta$, and therefore decreases for large $\tan\beta$.  

This distinction between the SM and MSSM frameworks has direct implications for the renormalization group (RG) evolution of neutrino parameters. Since the RG running of neutrino masses, mixing angles, and CP phases in the MSSM is dominantly governed by the tau Yukawa coupling $y_\tau$, its enhancement at large $\tan\beta$ can lead to substantial deviations from high-scale symmetry predictions (such as those imposed by $\mu$--$\tau$ reflection symmetry). Conversely, at small $\tan\beta$, the running effects are mild, and the low-energy values of neutrino parameters remain close to their high-scale inputs. While the SM Yukawa couplings are insensitive to $\tan\beta$, the MSSM framework introduces a strong $\tan\beta$ dependence, with large $\tan\beta$ amplifying RG effects and small $\tan\beta$ suppressing them. This makes $\tan\beta$ a key control parameter in assessing the phenomenological impact of RG evolution on the neutrino sector.  

At the SUSY breaking scale, appropriate matching conditions must be imposed to ensure the continuity of physical observables across the two regimes. These matching relations connect the parameters defined in the MSSM to their corresponding SM counterparts, allowing for a consistent transition between the two effective theories. This step is crucial to accurately capture the impact of supersymmetry on the low-energy neutrino observables and to quantify the radiative breaking effects of the $\mu$–$\tau$ reflection symmetry. The gauge couplings remain continuous across the SUSY threshold, while the Yukawa couplings require rescaling due to the presence of two Higgs doublets in the MSSM. The matching relations are given by
\begin{equation}
\begin{split}
        & g_i^{\text{MSSM}}(\Lambda_{\text{SUSY}}) = g_i^{\text{SM}}(\Lambda_{\text{SUSY}}), \\
        & y_t^{\text{MSSM}}(\Lambda_{\text{SUSY}}) = \frac{y_t^{\text{SM}}(\Lambda_{\text{SUSY}})}{\sin\beta}, \\
        & y_b^{\text{MSSM}}(\Lambda_{\text{SUSY}}) = \frac{y_b^{\text{SM}}(\Lambda_{\text{SUSY}})}{\cos\beta}, \\
        & y_\tau^{\text{MSSM}}(\Lambda_{\text{SUSY}}) = \frac{y_\tau^{\text{SM}}(\Lambda_{\text{SUSY}})}{\cos\beta}.
\end{split}
\label{matching}
\end{equation}

Here $g_i$ ($i=1,2,3$) denote the three gauge couplings, while $y_t$, $y_b$, and $y_\tau$ correspond to the top, bottom, and tau Yukawa couplings, respectively. These matching conditions explicitly highlight the role of $\tan\beta$ in relating the Yukawa couplings of the SM to those in the MSSM. In particular, the enhancement of $y_b$ and $y_\tau$ at large $\tan\beta$ is transmitted through these relations, thereby influencing the subsequent RG evolution of the neutrino parameters below the SUSY scale.

\section{Numerical results}
\begin{table}[t]
\centering
\begin{tabular}{c c c c c}
\hline
\textbf{Parameter} & \textbf{Input value at $m_t$ scale} & \multicolumn{3}{c}{\textbf{Output value at $\Lambda_{FS}$ for different $\tan\beta$}} \\ \cline{3-5}
              &            & \textbf{$tan\beta=10$} & \textbf{$tan\beta=30$} & \textbf{$tan\beta=58$} \\ \hline
$g_1$         & 0.46125    & 0.625793  &  0.625793  & 0.625793         \\ 
$g_2$         &0.66239     & 0.684943  & 0.684943   & 0.684943         \\ 
$g_3$         & 1.18955    & 0.722027   & 0.721910  & 0.721910         \\ 
$y_t$         & 0.9917     & 0.634446   & 0.640259 &  0.700936        \\ 
$y_b$         & 0.01571    & 0.065629  & 0.212362 & 0.601777         \\ 
$y_{\tau}$    & 0.01006    & 0.080029  & 0.259488  & 0.735524         \\ \hline

\end{tabular}
\caption{Input values of gauge and Yukawa coupling at $m_t=172GeV$ and corresponding output values at $\Lambda_{FS}=10^{14}GeV$ for three different values of $tan\beta$ with $\Lambda_s=7\ TeV$.} 
\label{gandy}
\end{table}

In this section, we aim to numerically investigate the breaking of the $\mu$–$\tau$ reflection symmetry by assuming its prediction of exact maximal values at a high-energy scale and examining the resulting deviations at the Electroweak scale chosen as the top quark mass scale. At low energy, the values of several neutrino parameters are constrained by global oscillation data. As presented in Table~\ref{GA}, we have the experimentally determined low energy values of the three mixing angles, the Dirac CP-violating phase $\delta$ and the two independent mass-squared differences. These are defined as $\Delta m_{21}^2 = m_2^2 - m_1^2$ and either $|\Delta m_{32}^2| = |m_3^2 - m_2^2|$ or $|\Delta m_{31}^2| = |m_3^2 - m_1^2|$, where $m_1$, $m_2$, and $m_3$ represent the three neutrino mass eigenvalues. Since oscillation experiments measure only the absolute values $|\Delta m_{32}^2|$ or $|\Delta m_{31}^2|$, two distinct mass orderings remain consistent with current data: the normal ordering (NO), characterized by $m_1 < m_2 < m_3$, and the inverted ordering (IO), with $m_3 < m_1 < m_2$. Furthermore, while oscillation experiments do not fix the absolute neutrino mass scale, cosmological observations impose an upper bound on the total neutrino mass, given by $\sum_i m_i < 0.12~\text{eV}$~\cite{mbound}. 

To perform the present analysis, we require the complete set of renormalization group equations (RGEs) along with the input values of the relevant parameters at the high-energy scale, which are subsequently evolved downward to obtain their corresponding values at the low-energy scale. In this study, we make use of the RGEs derived in our previous work \cite{CBorah} for twelve parameters — namely, the three neutrino mass eigenvalues, three mixing angles, and six CP-violating phases. These RGEs are coupled with those of the three gauge couplings and three Yukawa couplings. The RGEs for these coupling constants in both the Standard Model (SM) and the Minimal Supersymmetric Standard Model (MSSM) are well established in the literature and we have also presented them explicitly in our earlier work \cite{CBorah}. Similarly, the RGE for the Higgs quartic coupling in the SM scenario, which is also required in the analysis, is included there.

For the high-energy input values, we assume maximal values for the six CP phases and the Dirac phase $\delta$, as predicted by the $\mu$–$\tau$ reflection symmetry. The high-energy values of the coupling constants are obtained by evolving their experimentally measured low-energy counterparts upward using their respective RGEs, and are summarized in Table~\ref{gandy}. The remaining parameters, three mass eigenvalues and two mixing angles are treated as free parameters in our study. The primary challenge of our work is to identify the set of these five parameters that, when evolved, can successfully reproduce the low-energy experimental constraints in conjunction with the other parameters. Their optimal set of values is determined through a trial-and-error approach and is presented in the following tables.

After obtaining all parameter values at the high-energy scale $\Lambda_{FS}$, we proceed to solve the coupled differential equations in two stages. In the first stage, the equations are evolved from $\Lambda_{FS}$ down to the SUSY-breaking scale $\Lambda_{s}$, where the appropriate matching conditions are applied. In the second stage, the evolution continues from $\Lambda_{s}$ down to the electroweak scale $\Lambda_{EW}$. While choosing the mass eigenvalues at $\Lambda_{FS}$, we consider both the normal ordering (NO) and the inverted ordering (IO) scenarios. Furthermore, as mentioned earlier in section one, the $\mu$–$\tau$ reflection symmetry predicts two possible maximal values for the Dirac CP-violating phase $\delta$, leading to two corresponding choices for the six CP-violating phases (Eqs. \ref{ch1} and \ref{ch2}). Consequently, the analysis is performed separately for four distinct cases: NO–Case I, NO–Case II, IO–Case I, and IO–Case II.

\begin{figure}[!t]
  \centering
  \begin{tabular}{cc}
   \begin{subfigure}[b]{0.45\textwidth}
    \centering
    \includegraphics[height=5cm]{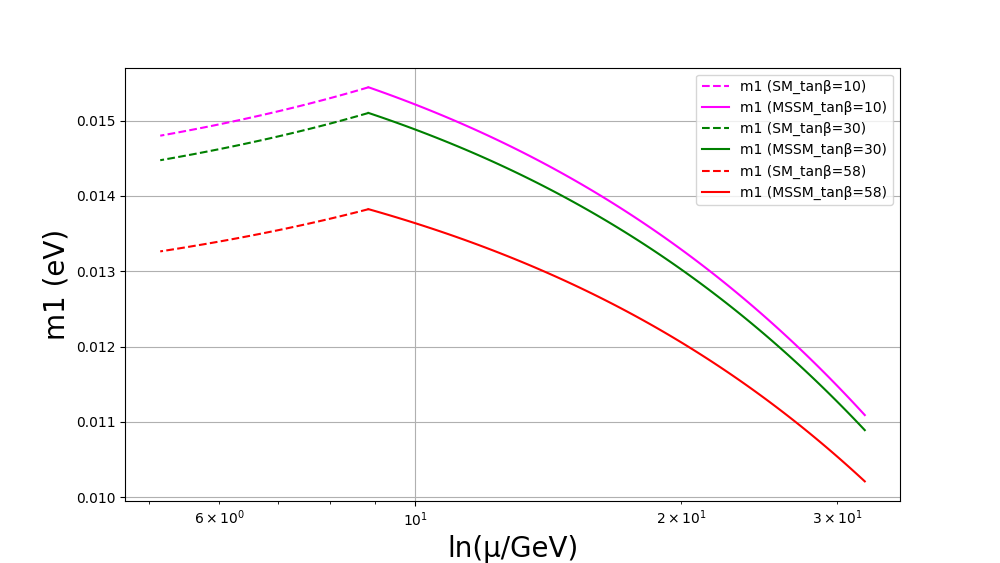}
    \caption{Evolution of $m_1$}
    \label{m1_no1}
  \end{subfigure}&
  \begin{subfigure}[b]{0.45\textwidth}
    \centering
    \includegraphics[height=5cm]{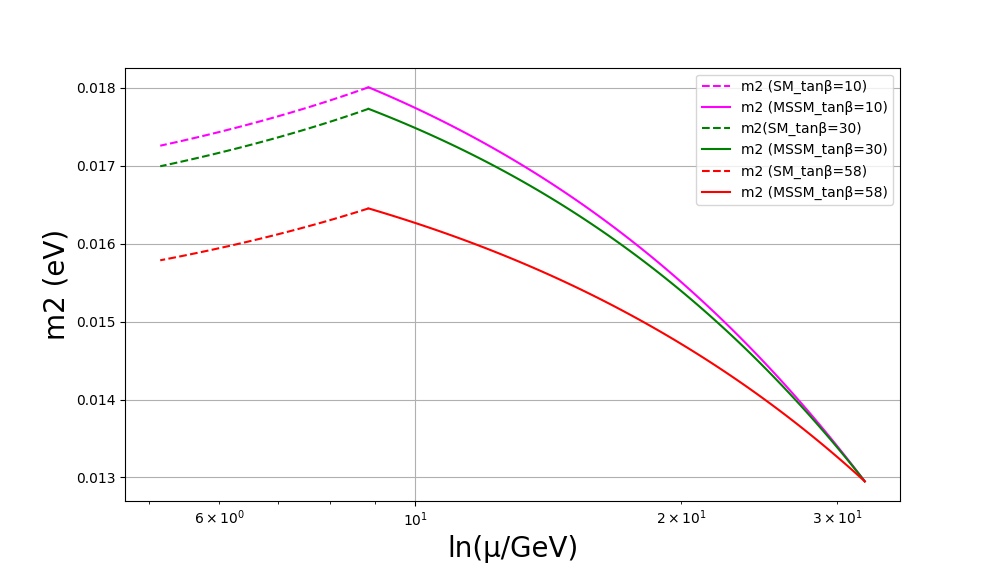}
    \caption{Evolution of $m_2$}
    \label{m2_no1}
  \end{subfigure} \\[2ex]
  \begin{subfigure}[b]{0.45\textwidth}
    \centering
    \includegraphics[height=5cm]{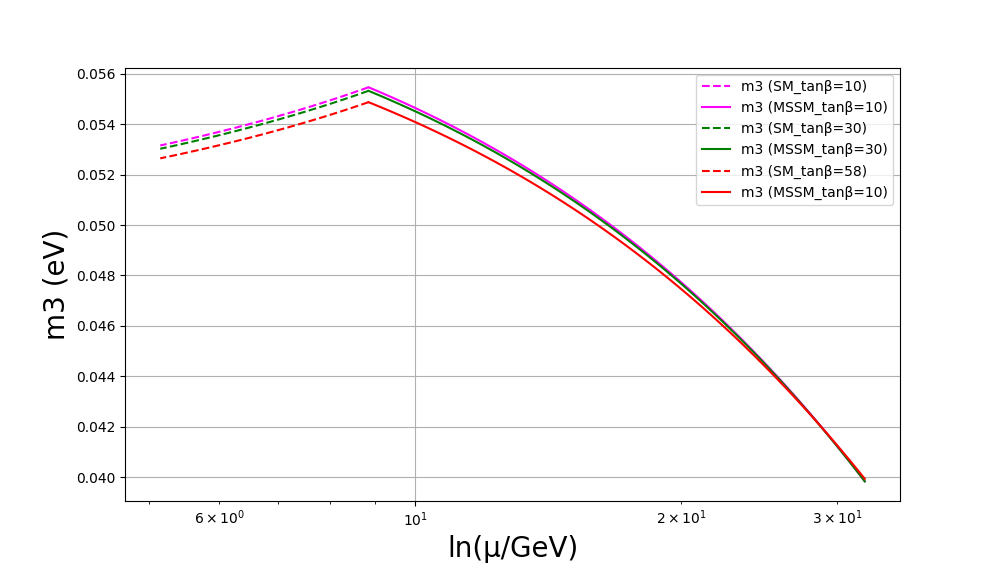}
    \caption{Evolution of $m_3$}
    \label{m3_no1}
  \end{subfigure} &
  \begin{subfigure}[b]{0.45\textwidth}
    \centering
    \includegraphics[height=5cm]{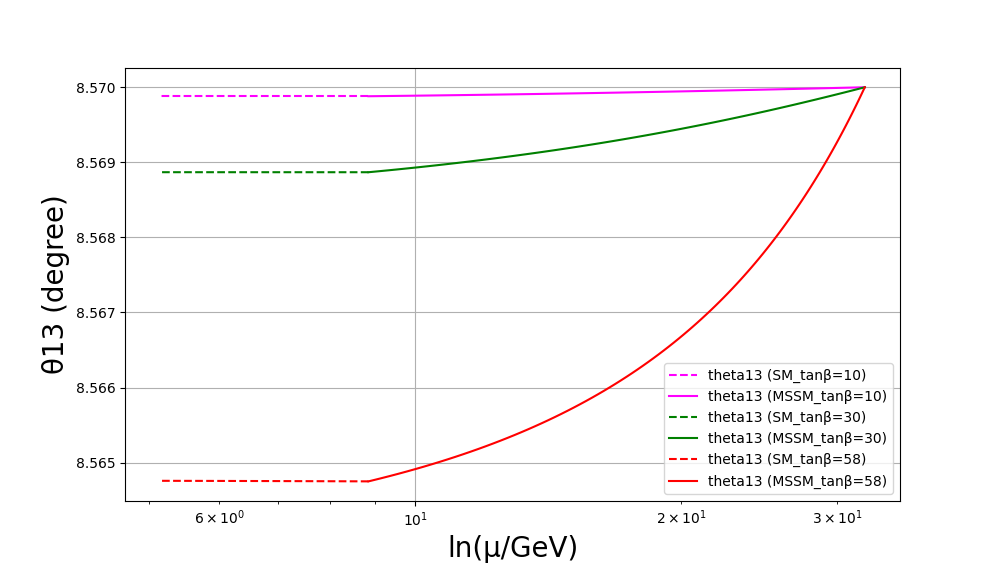}
    \caption{Evolution of $\theta_{13}$}
    \label{theta12_no1}
  \end{subfigure} \\[2ex]
  \begin{subfigure}[b]{0.45\textwidth}
    \centering
    \includegraphics[height=5cm]{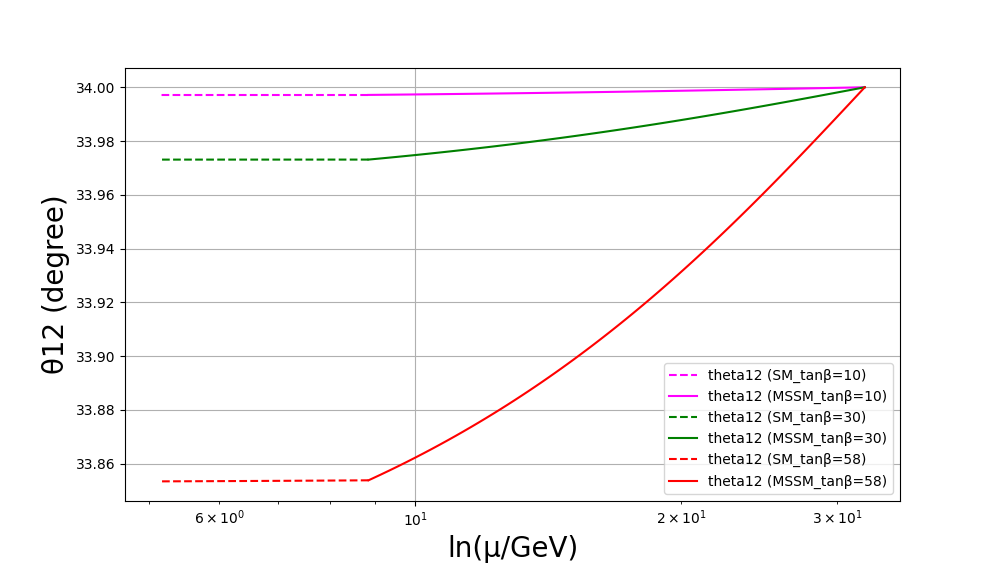}
    \caption{Evolution of $\theta_{12}$}
    \label{theta13_no1}
  \end{subfigure} &
  \begin{subfigure}[b]{0.45\textwidth}
    \centering
    \includegraphics[height=5cm]{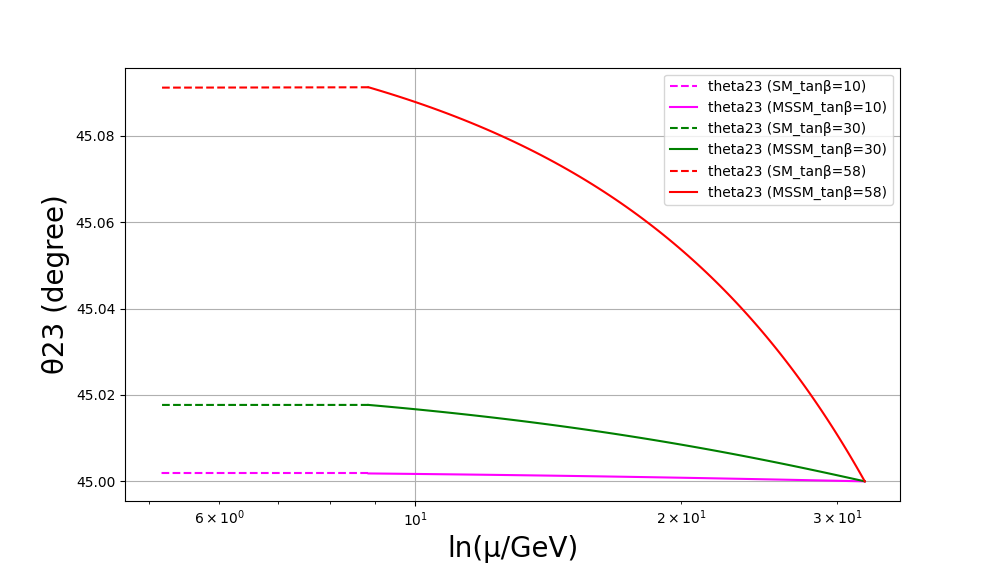}
    \caption{Evolution of $\theta_{23}$}
    \label{theta23_no1}
  \end{subfigure} \\[2ex]
    \end{tabular}
  \caption{Evolution of mass eigenvalues and mixing angles with energy scale for NO and case-I with three different values of $\tan\beta$. Solid and dashed portions of each curve represent the evolution in the MSSM and SM regions respectively. Magenta , green and red colors respectively stands for $\tan\beta=10,\  30\ \text{and}\ 58$.}
  \label{FNO11}
\end{figure}

\begin{figure}[!t]
  \centering
  \begin{tabular}{cc}
  \begin{subfigure}[b]{0.45\textwidth}
    \centering
    \includegraphics[height=5cm]{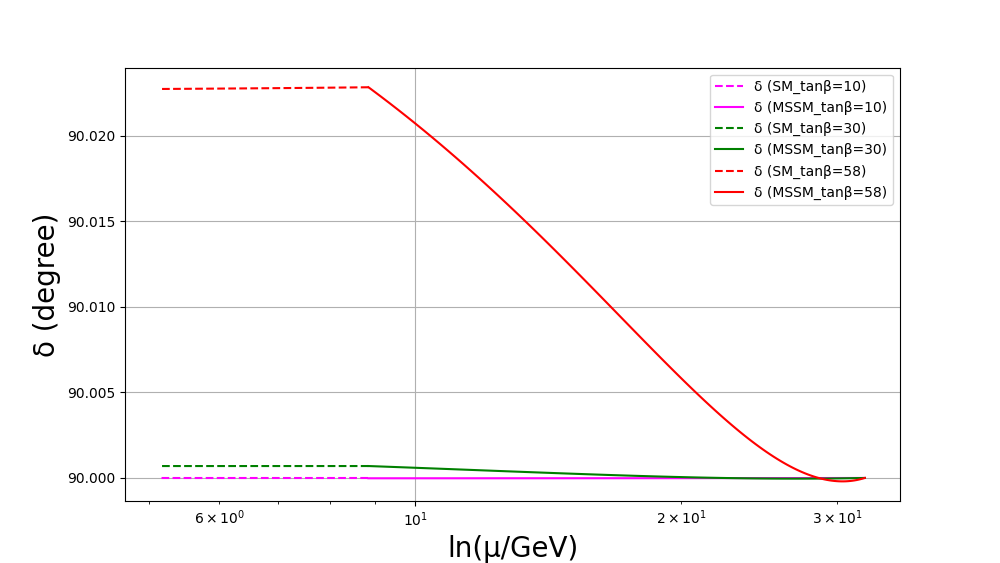}
    \caption{Evolution of $\delta$}
    \label{delta_no1}
  \end{subfigure} &
  \begin{subfigure}[b]{0.45\textwidth}
    \centering
    \includegraphics[height=5cm]{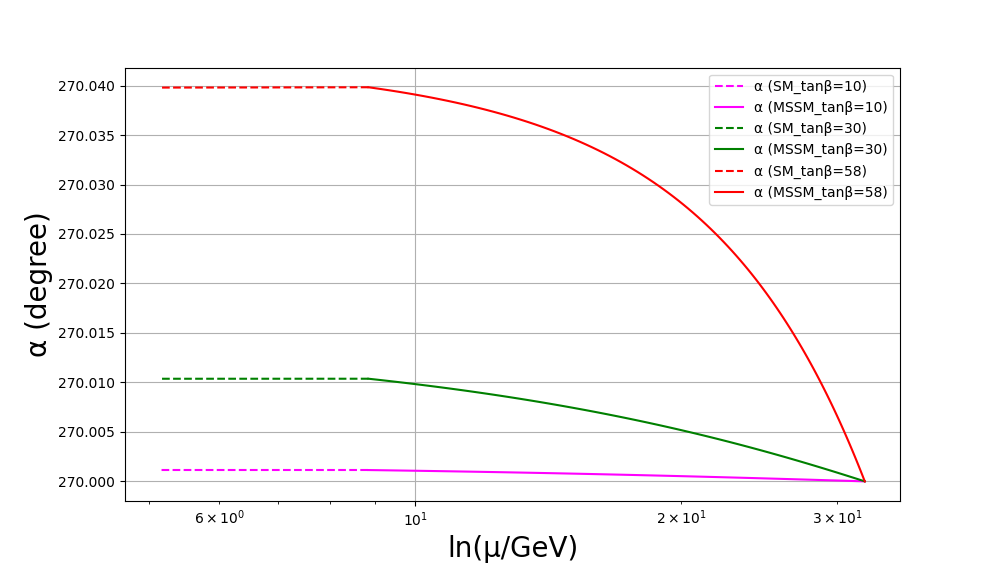}
    \caption{Evolution of $\alpha$}
  \end{subfigure} \\[2ex]
  \begin{subfigure}[b]{0.45\textwidth}
    \centering
    \includegraphics[height=5cm]{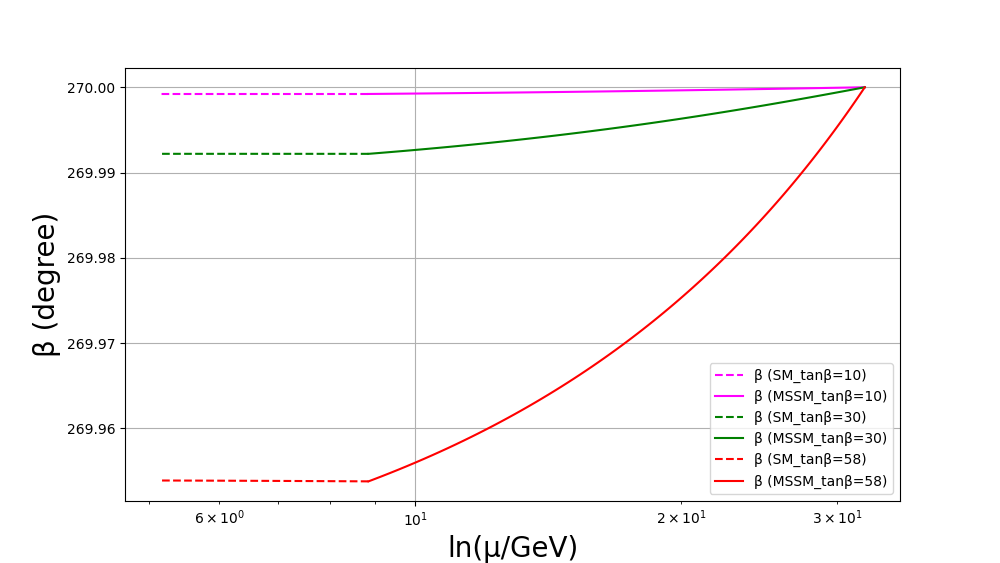}
    \caption{Evolution of $\beta$ }
  \end{subfigure} &
  \begin{subfigure}[b]{0.45\textwidth}
    \centering
    \includegraphics[height=5cm]{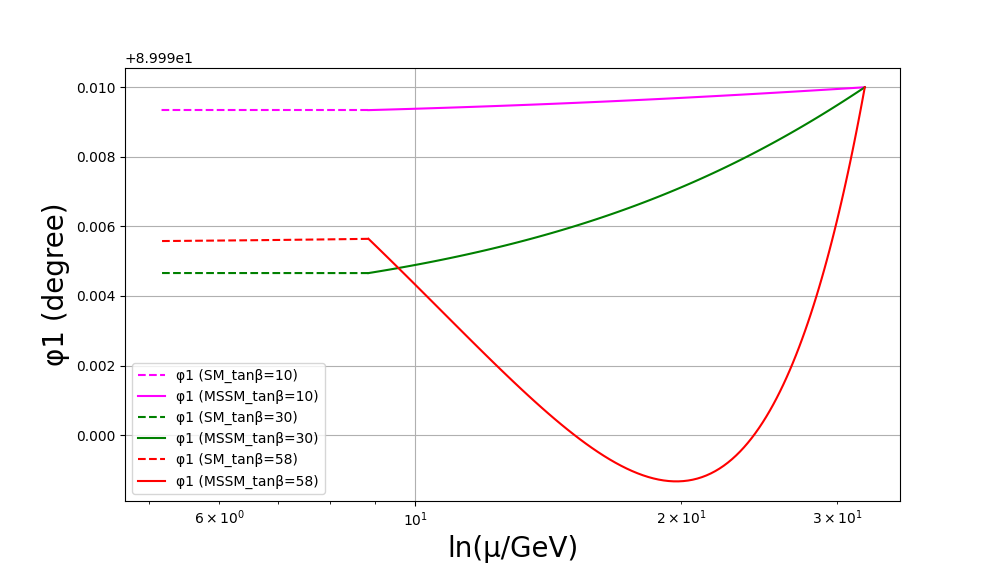}
    \caption{Evolution of $\phi_1$}
  \end{subfigure} \\[2ex]
  \begin{subfigure}[b]{0.45\textwidth}
    \centering
    \includegraphics[height=5cm]{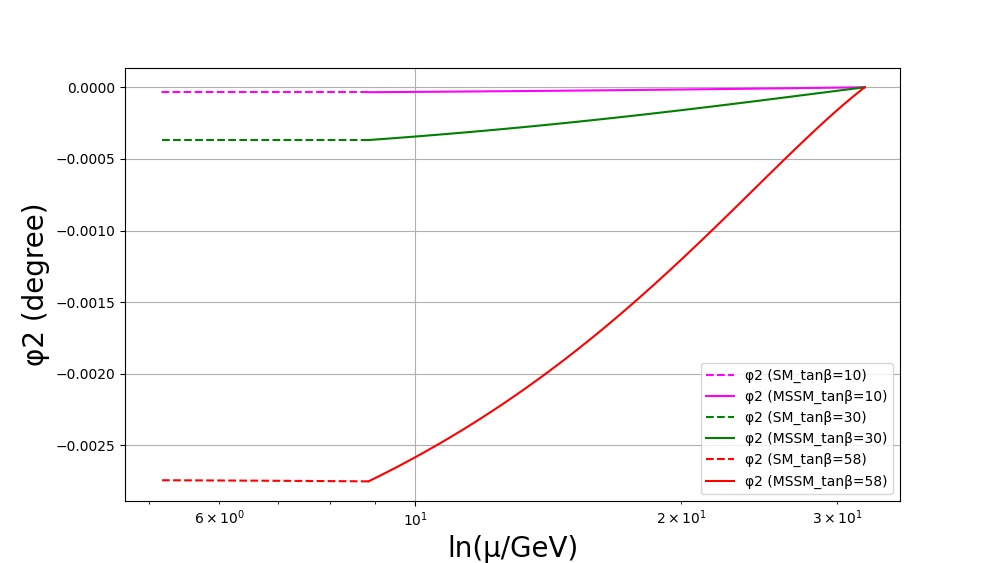}
    \caption{Evolution of $\phi_2$}
  \end{subfigure} & 
  \begin{subfigure}[b]{0.45\textwidth}
    \centering
    \includegraphics[height=5cm]{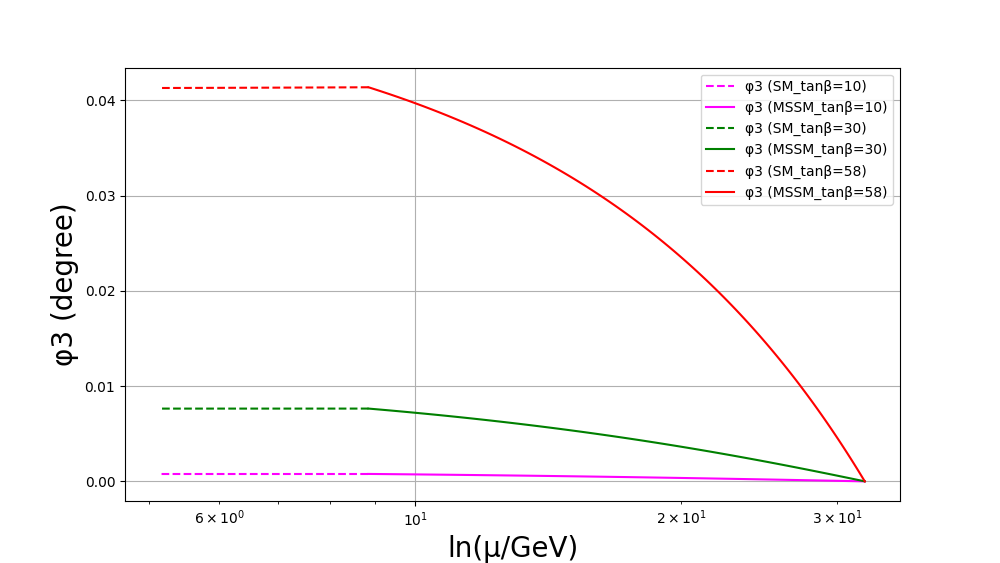}
    \caption{Evolution of $\phi_3$}
    \label{phi3_no1}
  \end{subfigure}  
  \end{tabular}
  \caption{Evolution of CP phases with energy scale for NO and case-I with three different values of $\tan\beta$. Solid and dashed portions of each curve represent the evolution in the MSSM and SM regions respectively. Magenta , green and red colors respectively stands for $\tan\beta=10,\  30\ \text{and}\ 58$.}
  \label{FNO12}
\end{figure}

\begin{figure}[!t]
  \centering
  \begin{tabular}{cc}
   \begin{subfigure}[b]{0.45\textwidth}
    \centering
    \includegraphics[height=5cm]{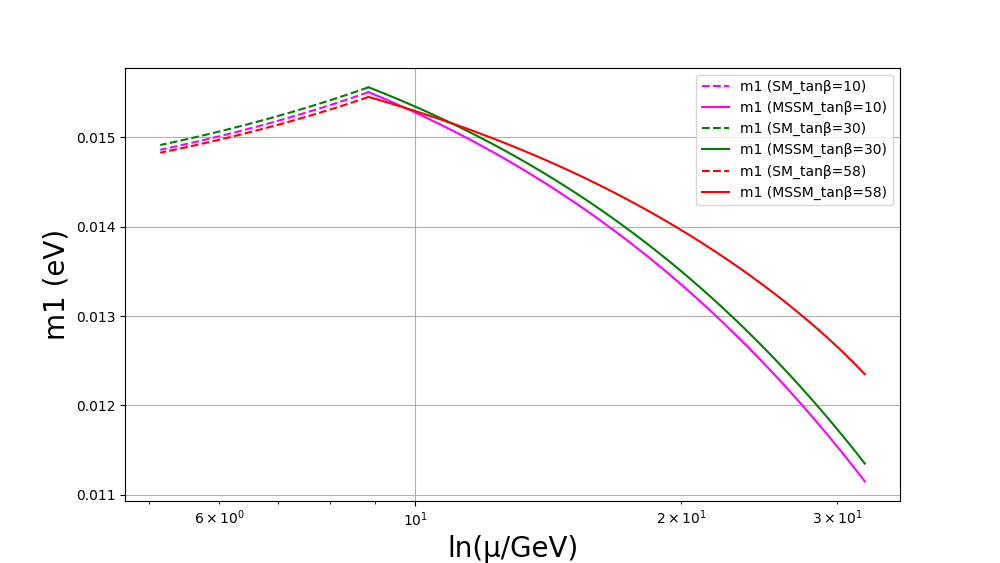}
    \caption{Evolution of $m_1$}
    \label{m1_no1}
  \end{subfigure}&
  \begin{subfigure}[b]{0.45\textwidth}
    \centering
    \includegraphics[height=5cm]{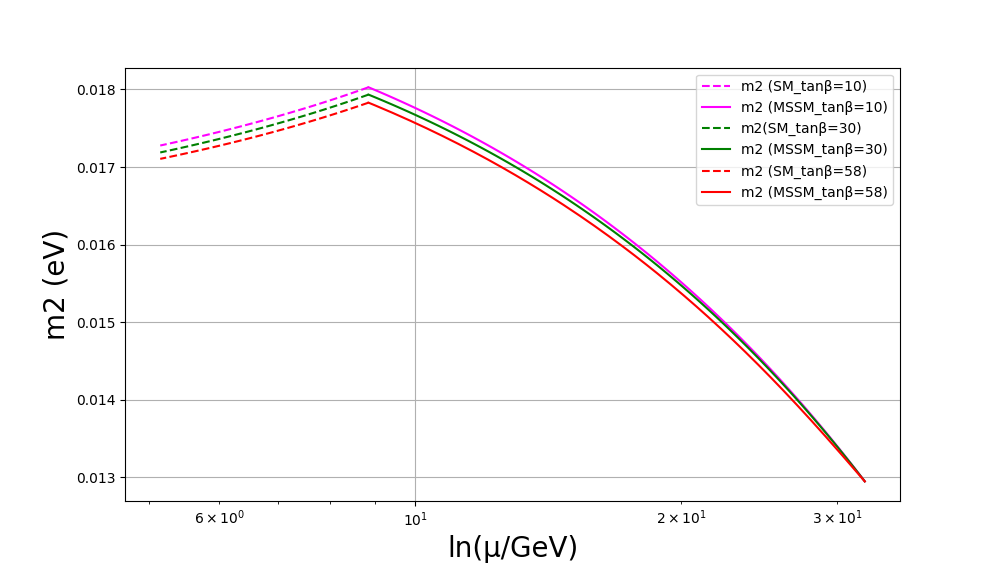}
    \caption{Evolution of $m_2$}
    \label{m2_no1}
  \end{subfigure} \\[2ex]
  \begin{subfigure}[b]{0.45\textwidth}
    \centering
    \includegraphics[height=5cm]{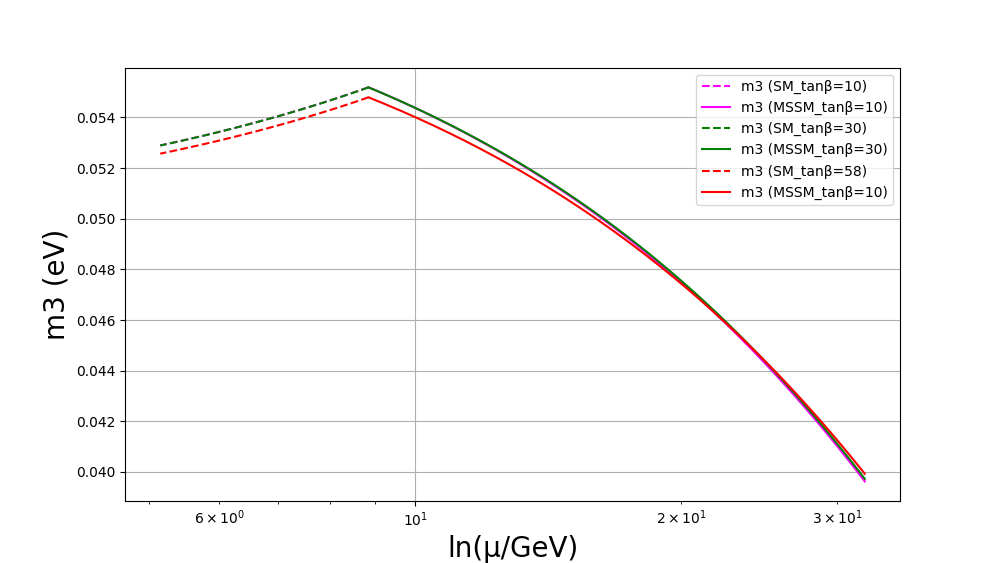}
    \caption{Evolution of $m_3$}
    \label{m3_no1}
  \end{subfigure} &
  \begin{subfigure}[b]{0.45\textwidth}
    \centering
    \includegraphics[height=5cm]{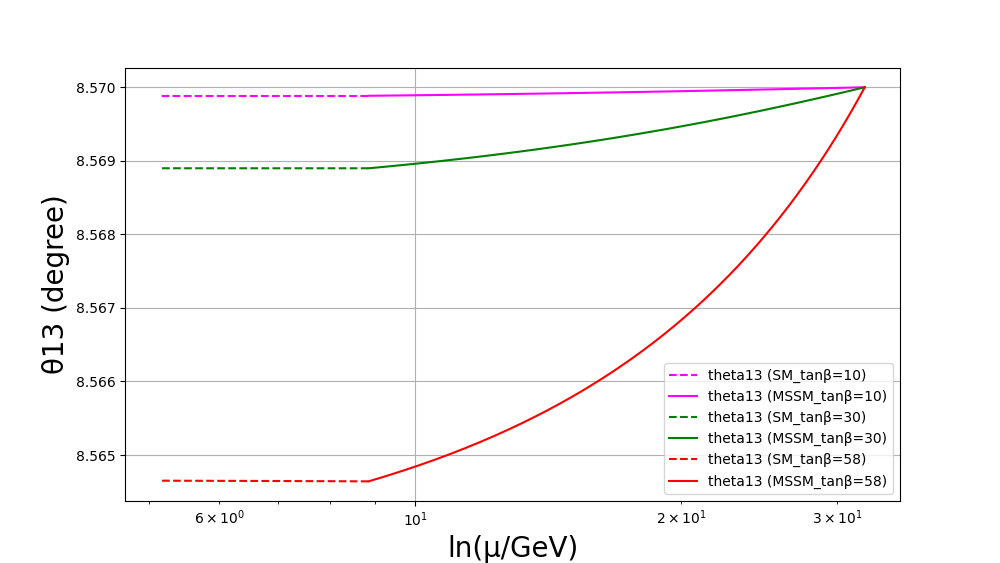}
    \caption{Evolution of $\theta_{13}$}
    \label{theta12_no1}
  \end{subfigure} \\[2ex]
  \begin{subfigure}[b]{0.45\textwidth}
    \centering
    \includegraphics[height=5cm]{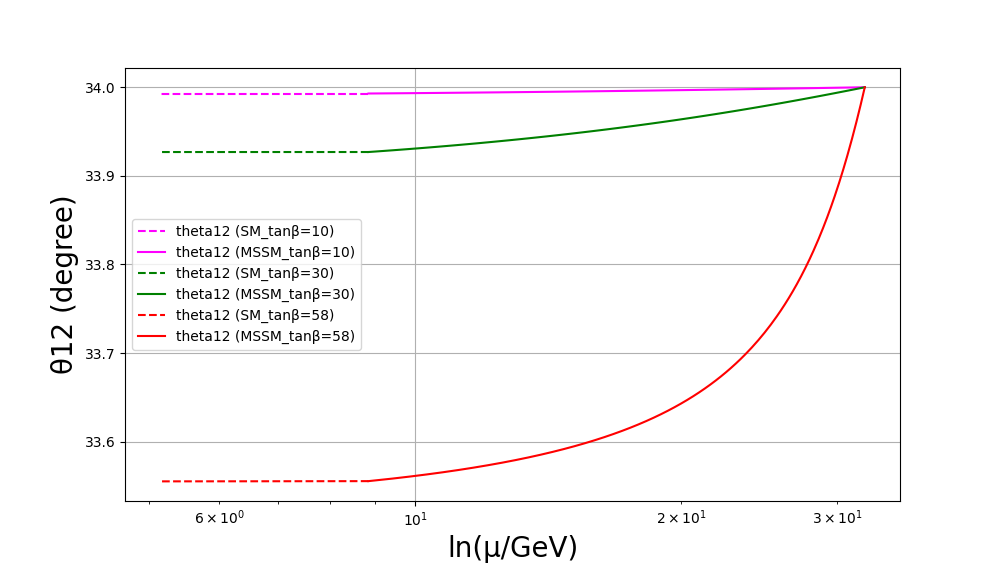}
    \caption{Evolution of $\theta_{12}$}
    \label{theta13_no1}
  \end{subfigure} &
  \begin{subfigure}[b]{0.45\textwidth}
    \centering
    \includegraphics[height=5cm]{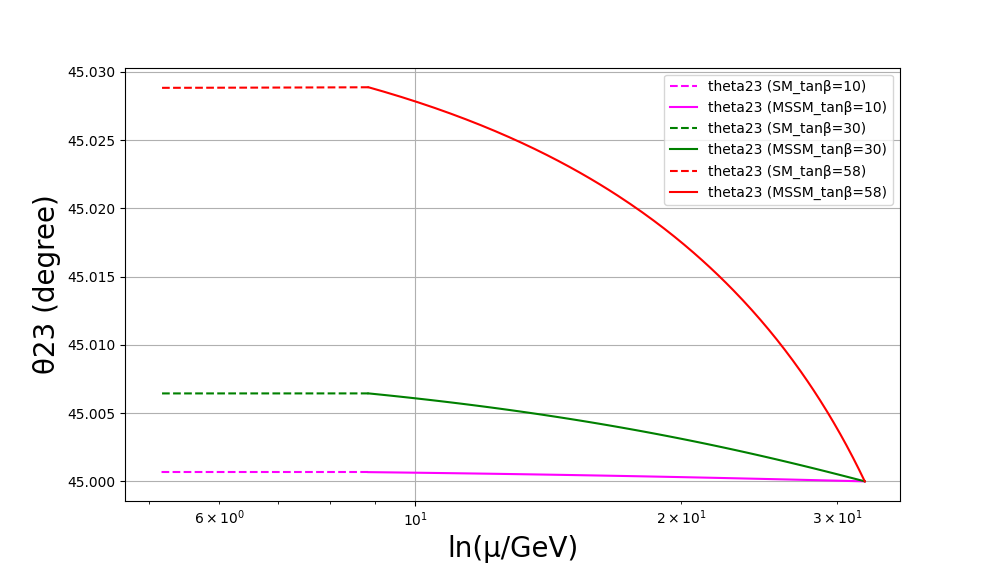}
    \caption{Evolution of $\theta_{23}$}
    \label{theta23_no1}
  \end{subfigure} \\[2ex]
    \end{tabular}
  \caption{Evolution of mass eigenvalues and mixing angles with energy scale for NO and case-II with three different values of $\tan\beta$. Solid and dashed portions of each curve represent the evolution in the MSSM and SM regions respectively. Magenta, green and red colors respectively stand for $\tan\beta=10,\  30\ \text{and}\ 58$.}
  \label{FNO21}
\end{figure}

\begin{figure}[!t]
  \centering
  \begin{tabular}{cc}
  \begin{subfigure}[b]{0.45\textwidth}
    \centering
    \includegraphics[height=5cm]{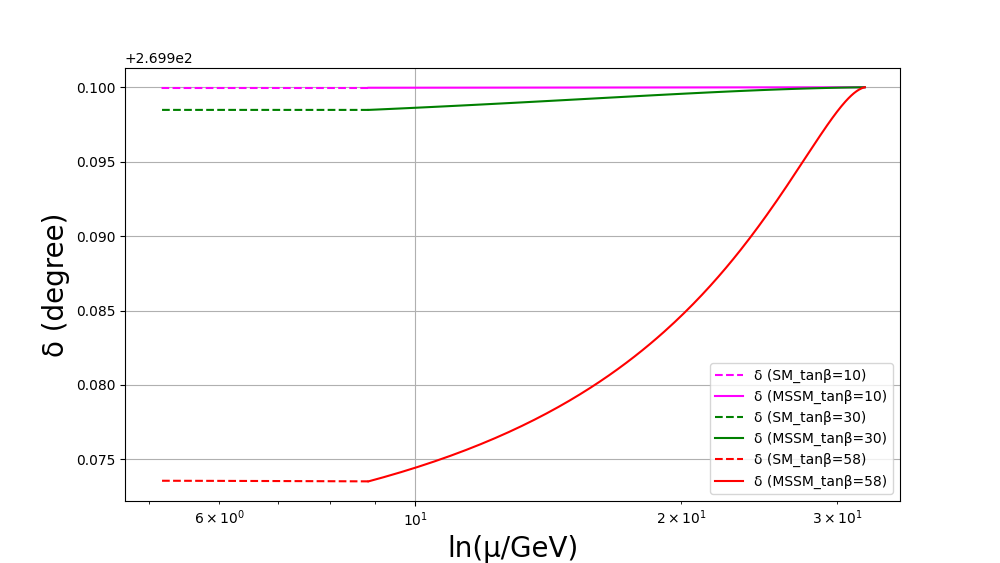}
    \caption{Evolution of $\delta$}
    \label{delta_no1}
  \end{subfigure} &
  \begin{subfigure}[b]{0.45\textwidth}
    \centering
    \includegraphics[height=5cm]{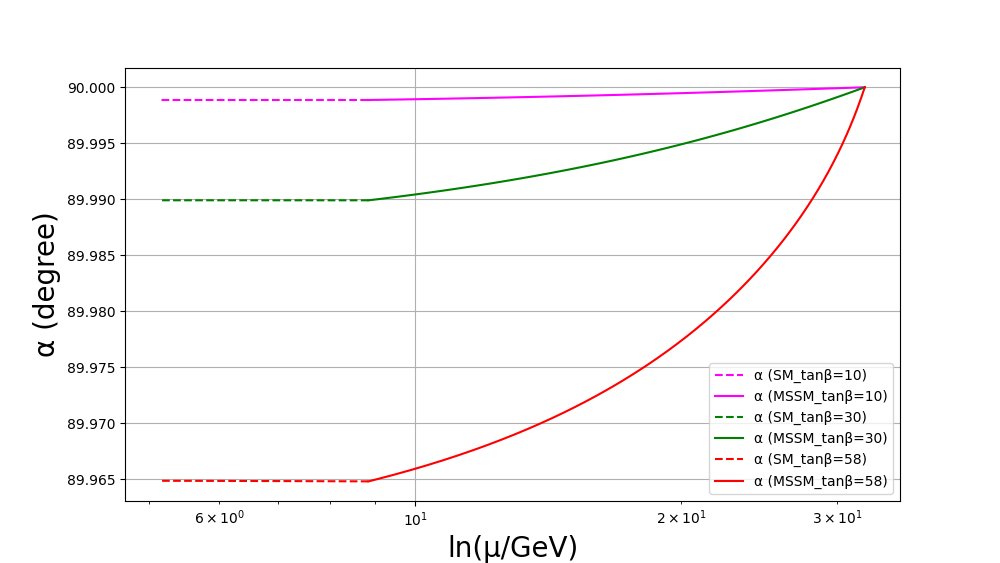}
    \caption{Evolution of $\alpha$}
  \end{subfigure} \\[2ex]
  \begin{subfigure}[b]{0.45\textwidth}
    \centering
    \includegraphics[height=5cm]{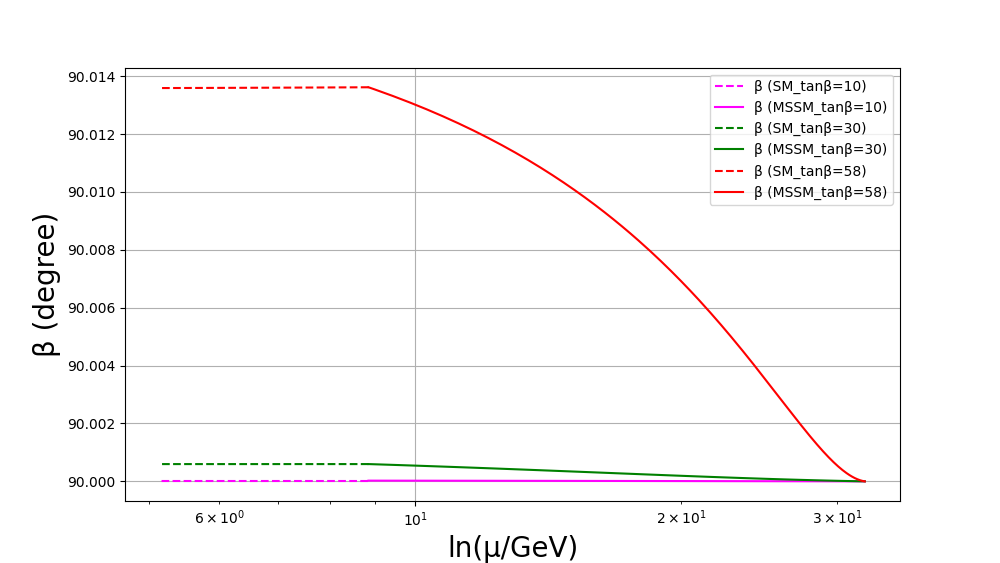}
    \caption{Evolution of $\beta$ }
  \end{subfigure} &
  \begin{subfigure}[b]{0.45\textwidth}
    \centering
    \includegraphics[height=5cm]{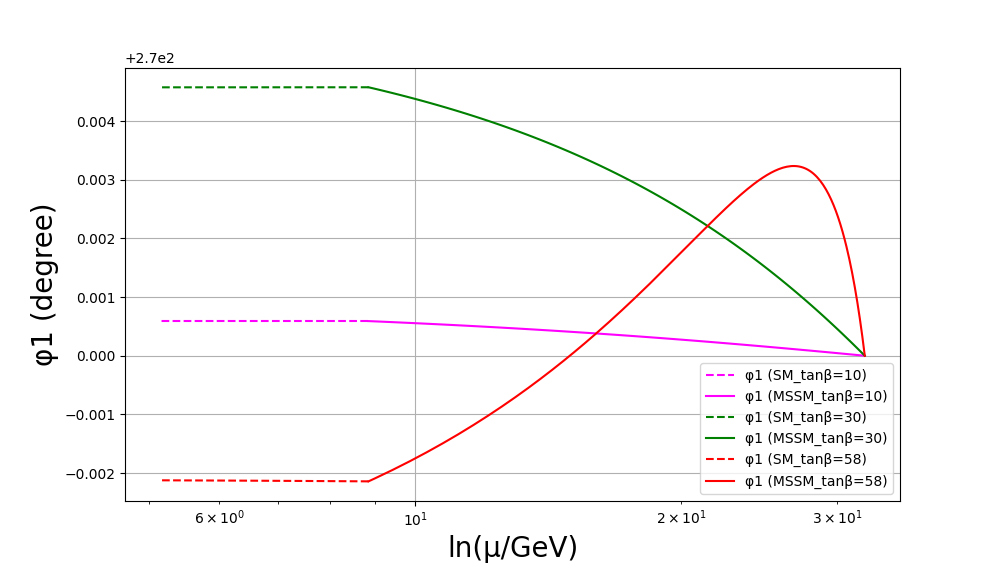}
    \caption{Evolution of $\phi_1$}
  \end{subfigure} \\[2ex]
  \begin{subfigure}[b]{0.45\textwidth}
    \centering
    \includegraphics[height=5cm]{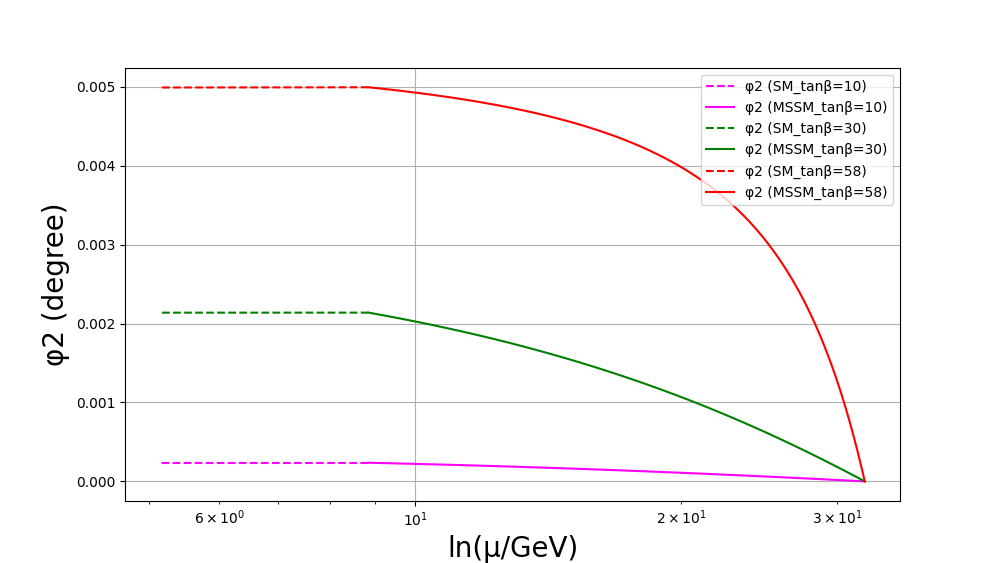}
    \caption{Evolution of $\phi_2$}
  \end{subfigure} & 
  \begin{subfigure}[b]{0.45\textwidth}
    \centering
    \includegraphics[height=5cm]{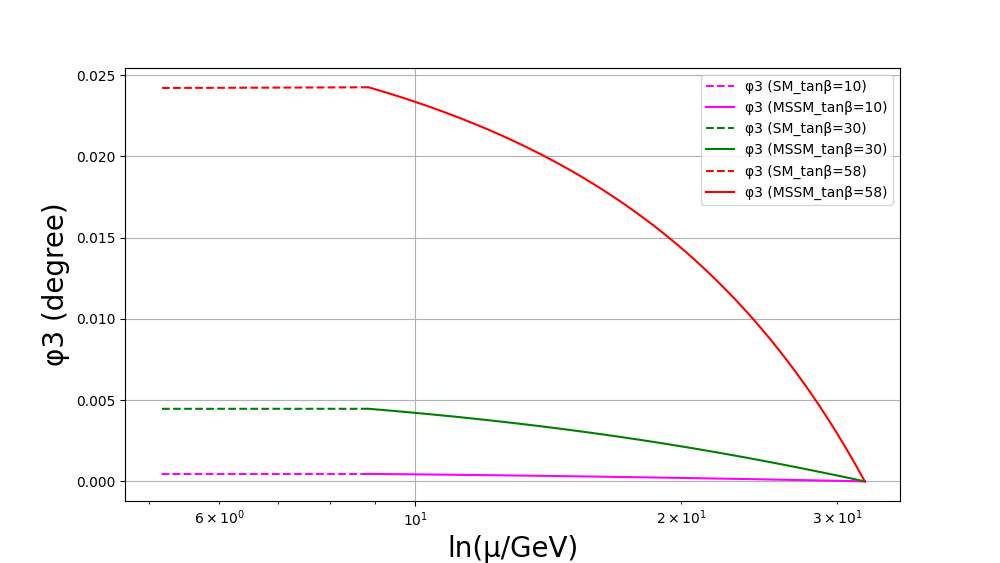}
    \caption{Evolution of $\phi_3$}
    \label{phi3_no1}
  \end{subfigure}  
  \end{tabular}
  \caption{Evolution of CP phases with energy scale for NO and case-II with three different values of $\tan\beta$. Solid and dashed portions of each curve represent the evolution in the MSSM and SM regions respectively. Magenta , green and red colors respectively stands for $\tan\beta=10,\  30\ \text{and}\ 58$.}
  \label{FNO22}
\end{figure}

\begin{figure}[!t]
  \centering
  \begin{tabular}{cc}
   \begin{subfigure}[b]{0.45\textwidth}
    \centering
    \includegraphics[height=5cm]{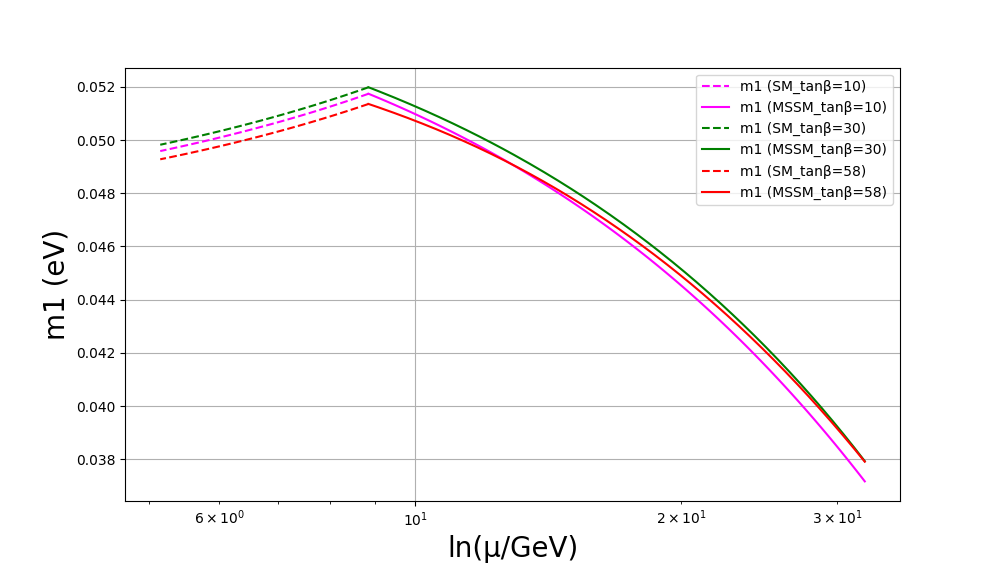}
    \caption{Evolution of $m_1$}
    \label{m1_no1}
  \end{subfigure}&
  \begin{subfigure}[b]{0.45\textwidth}
    \centering
    \includegraphics[height=5cm]{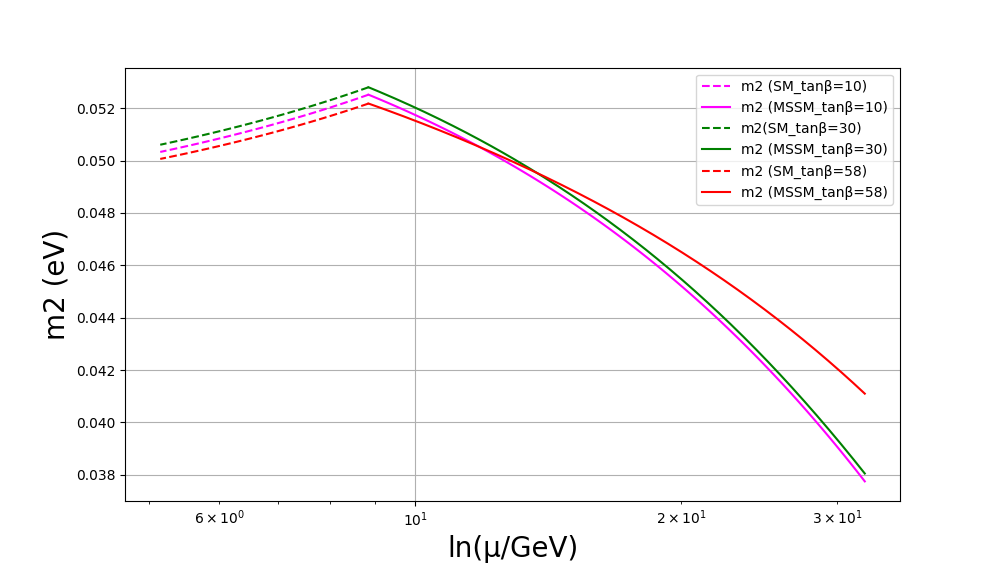}
    \caption{Evolution of $m_2$}
    \label{m2_no1}
  \end{subfigure} \\[2ex]
  \begin{subfigure}[b]{0.45\textwidth}
    \centering
    \includegraphics[height=5cm]{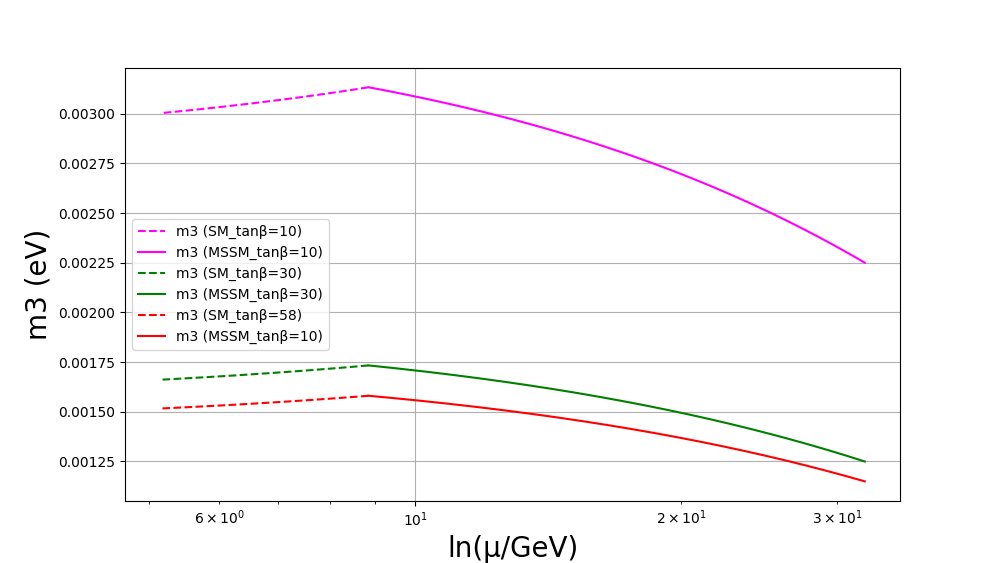}
    \caption{Evolution of $m_3$}
    \label{m3_no1}
  \end{subfigure} &
  \begin{subfigure}[b]{0.45\textwidth}
    \centering
    \includegraphics[height=5cm]{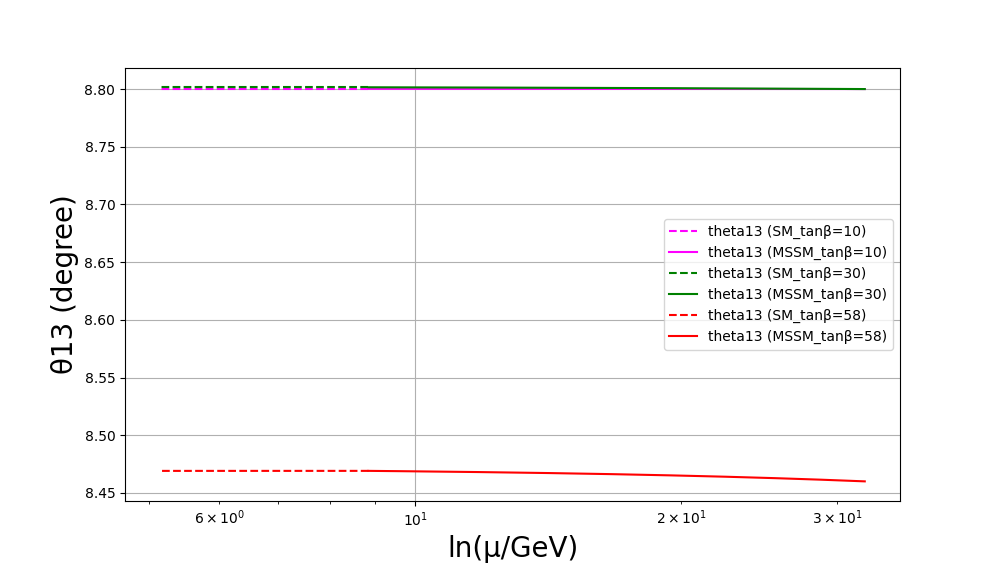}
    \caption{Evolution of $\theta_{13}$}
    \label{theta12_no1}
  \end{subfigure} \\[2ex]
  \begin{subfigure}[b]{0.45\textwidth}
    \centering
    \includegraphics[height=5cm]{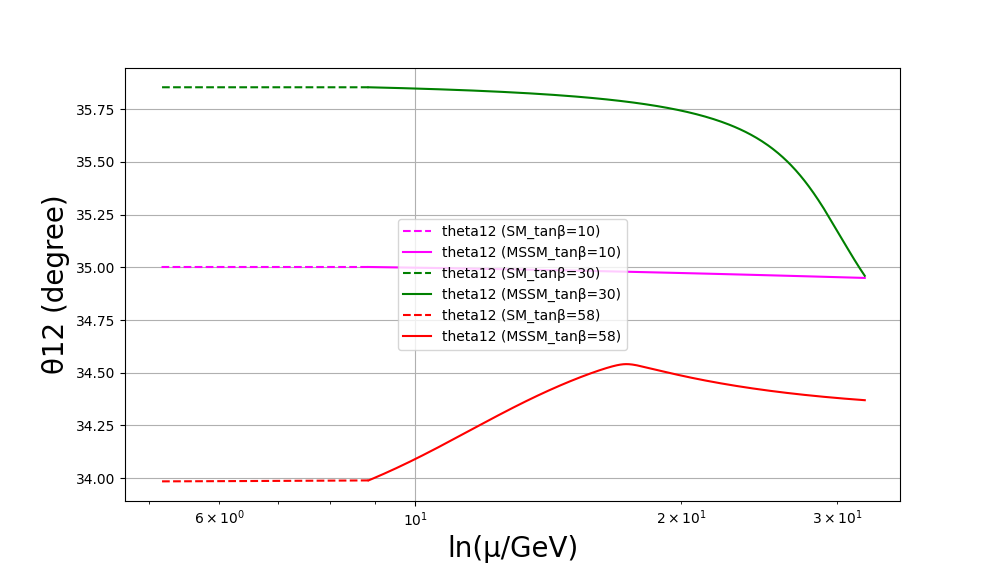}
    \caption{Evolution of $\theta_{12}$}
    \label{theta13_no1}
  \end{subfigure} &
  \begin{subfigure}[b]{0.45\textwidth}
    \centering
    \includegraphics[height=5cm]{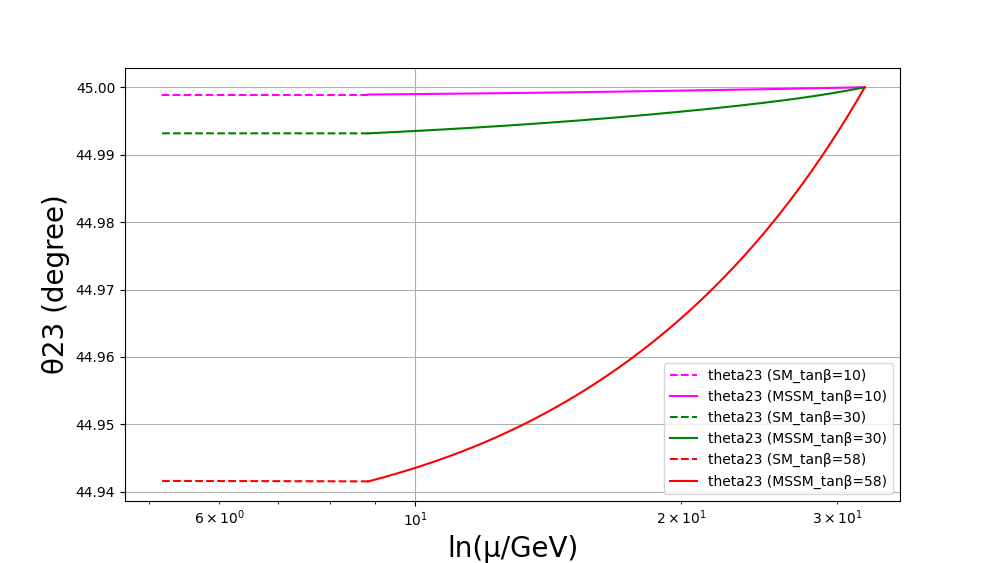}
    \caption{Evolution of $\theta_{23}$}
    \label{theta23_no1}
  \end{subfigure} \\[2ex]
    \end{tabular}
  \caption{Evolution of mass eigenvalues and mixing angles with energy scale for IO and case-I with three different values of $\tan\beta$. Solid and dashed portions of each curve represent the evolution in the MSSM and SM regions respectively. Magenta , green and red colors respectively stands for $\tan\beta=10,\  30\ \text{and}\ 58$.}
  \label{FIO11}
\end{figure}

\begin{figure}[!t]
  \centering
  \begin{tabular}{cc}
  \begin{subfigure}[b]{0.45\textwidth}
    \centering
    \includegraphics[height=5cm]{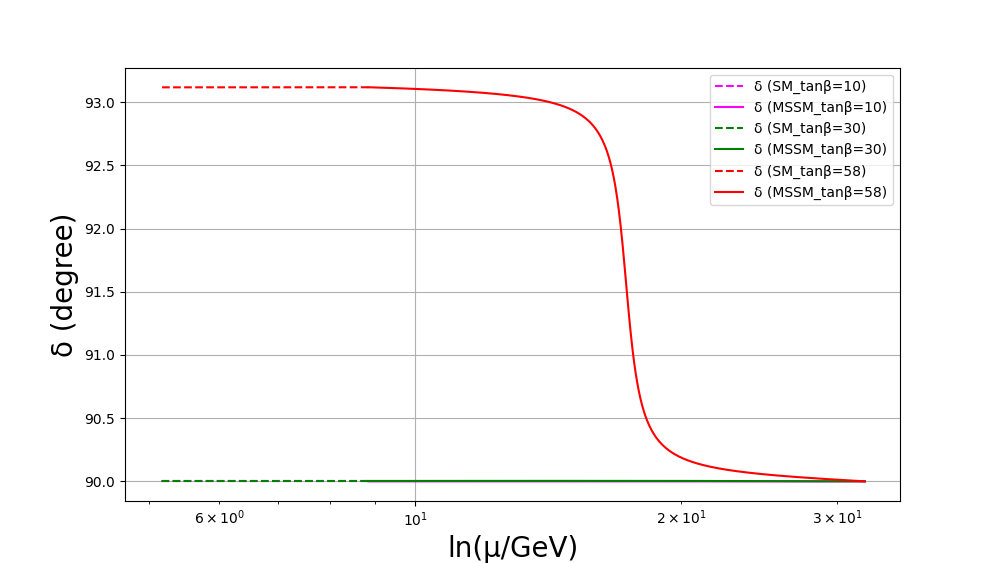}
    \caption{Evolution of $\delta$}
    \label{delta_no1}
  \end{subfigure} &
  \begin{subfigure}[b]{0.45\textwidth}
    \centering
    \includegraphics[height=5cm]{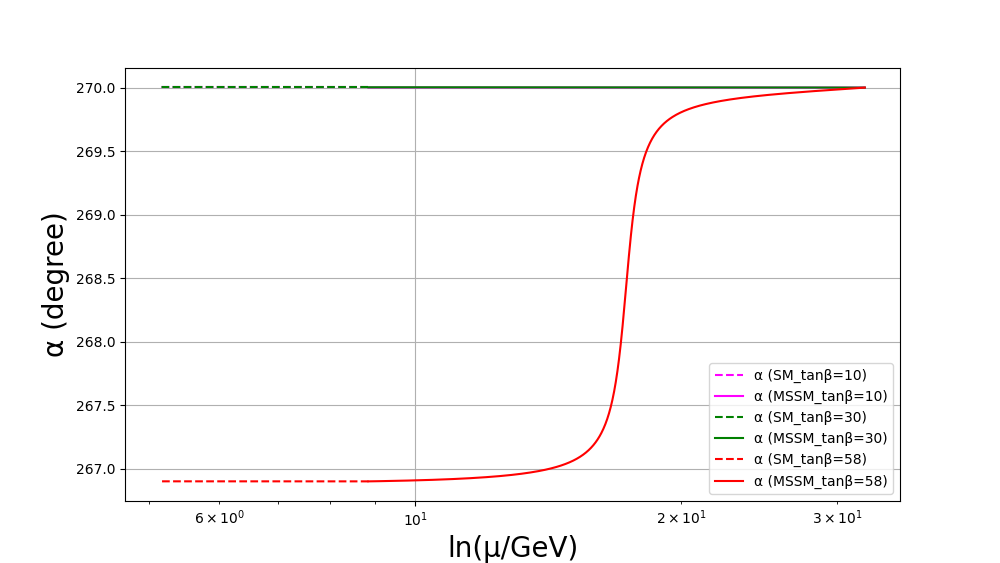}
    \caption{Evolution of $\alpha$}
  \end{subfigure} \\[2ex]
  \begin{subfigure}[b]{0.45\textwidth}
    \centering
    \includegraphics[height=5cm]{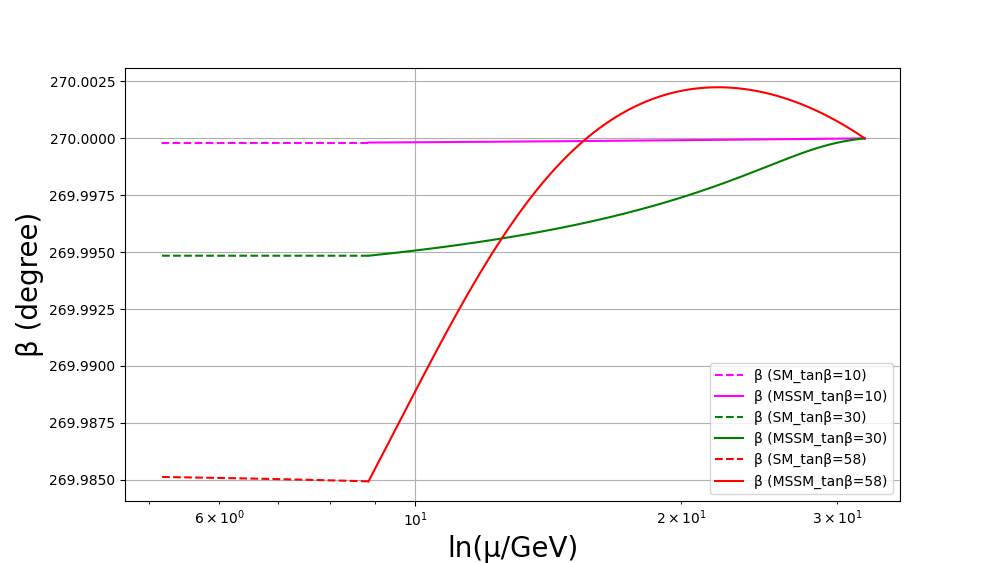}
    \caption{Evolution of $\beta$ }
  \end{subfigure} &
  \begin{subfigure}[b]{0.45\textwidth}
    \centering
    \includegraphics[height=5cm]{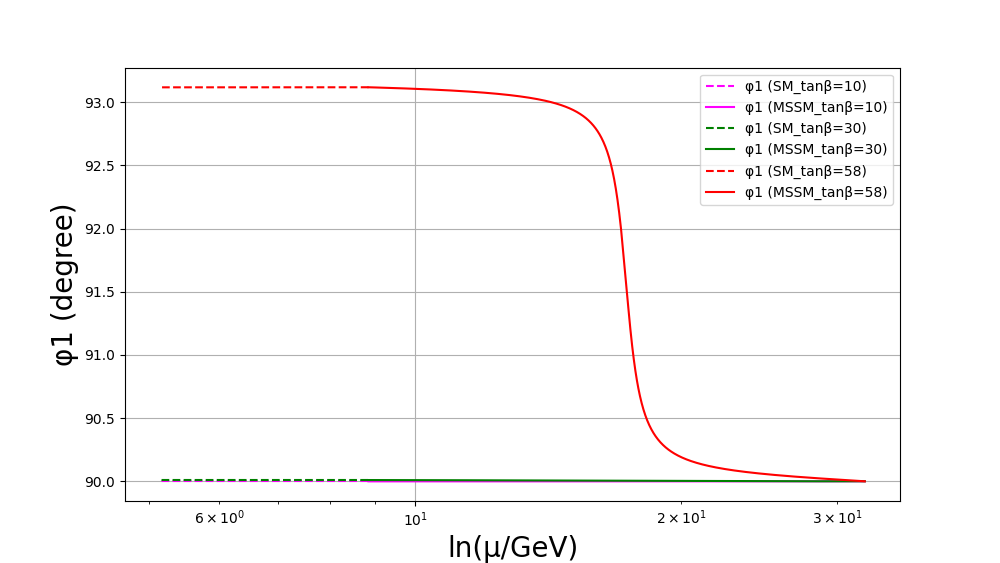}
    \caption{Evolution of $\phi_1$}
  \end{subfigure} \\[2ex]
  \begin{subfigure}[b]{0.45\textwidth}
    \centering
    \includegraphics[height=5cm]{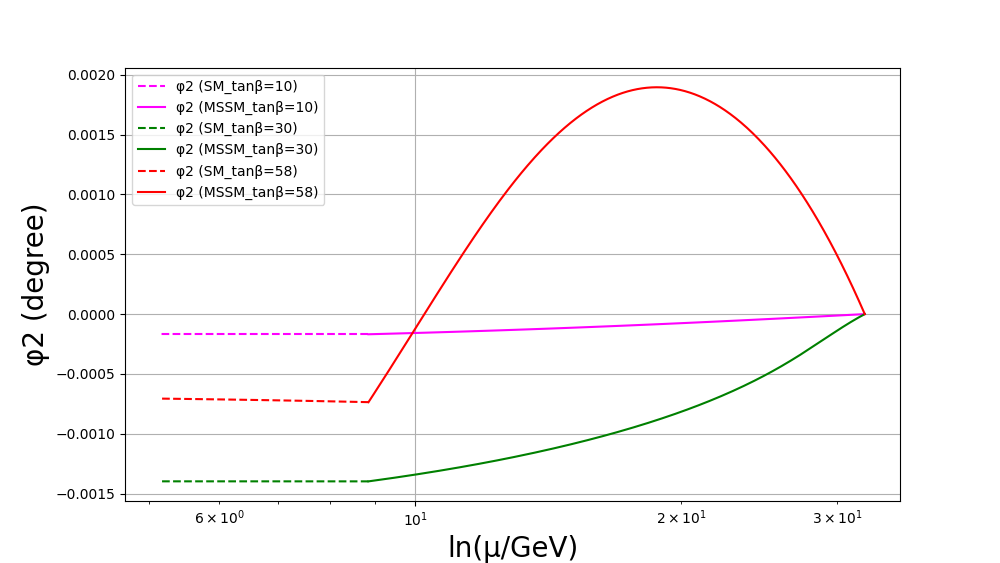}
    \caption{Evolution of $\phi_2$}
  \end{subfigure} & 
  \begin{subfigure}[b]{0.45\textwidth}
    \centering
    \includegraphics[height=5cm]{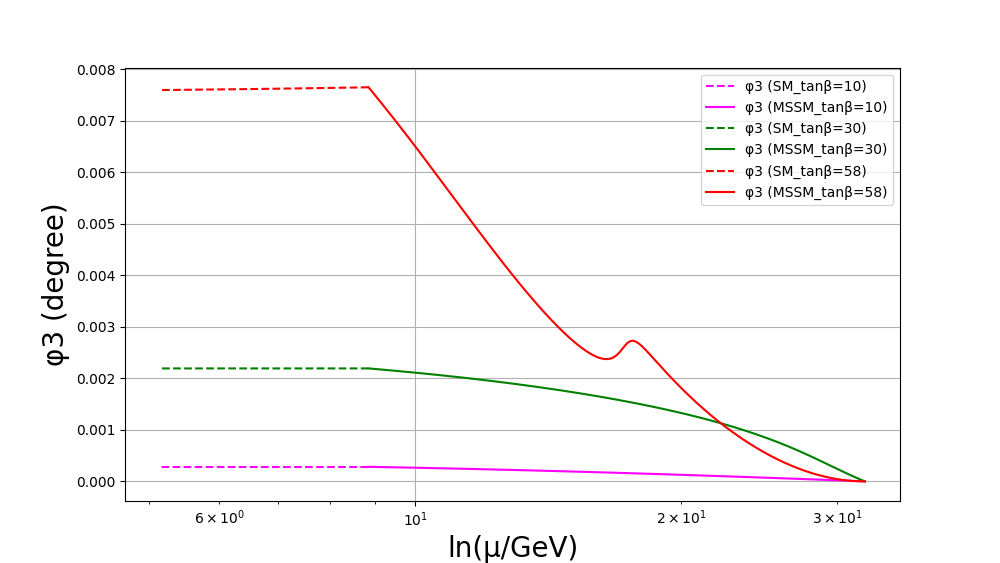}
    \caption{Evolution of $\phi_3$}
    \label{phi3_no1}
  \end{subfigure}  
  \end{tabular}
  \caption{Evolution of CP phases with energy scale for IO and case-I with three different values of $\tan\beta$. Solid and dashed portions of each curve represent the evolution in the MSSM and SM regions respectively. Magenta, green and red colors respectively stands for $\tan\beta=10,\  30\ \text{and}\ 58$.}
  \label{FIO12}
\end{figure}

\begin{figure}[!t]
  \centering
  \begin{tabular}{cc}
   \begin{subfigure}[b]{0.45\textwidth}
    \centering
    \includegraphics[height=5cm]{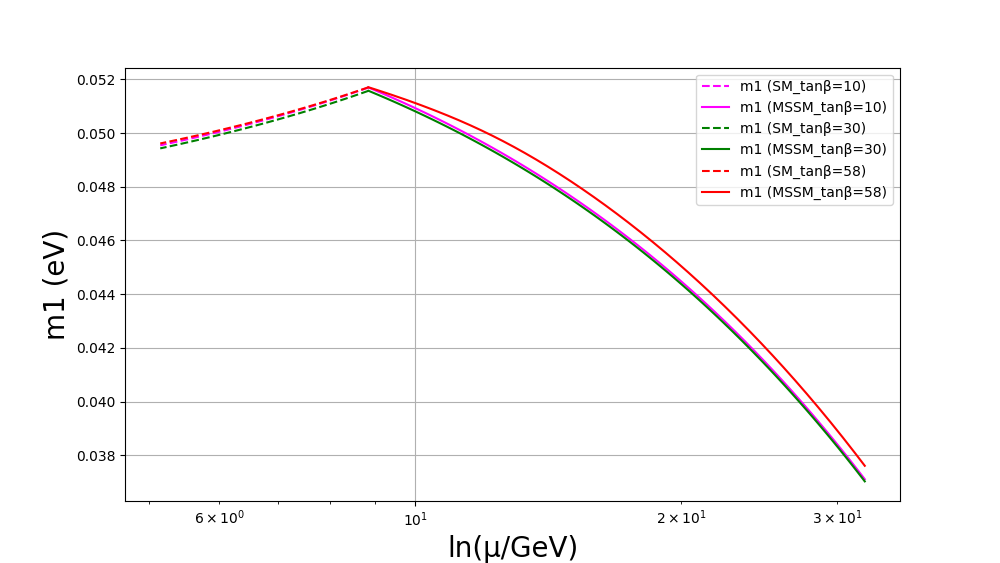}
    \caption{Evolution of $m_1$}
    \label{m1_no1}
  \end{subfigure}&
  \begin{subfigure}[b]{0.45\textwidth}
    \centering
    \includegraphics[height=5cm]{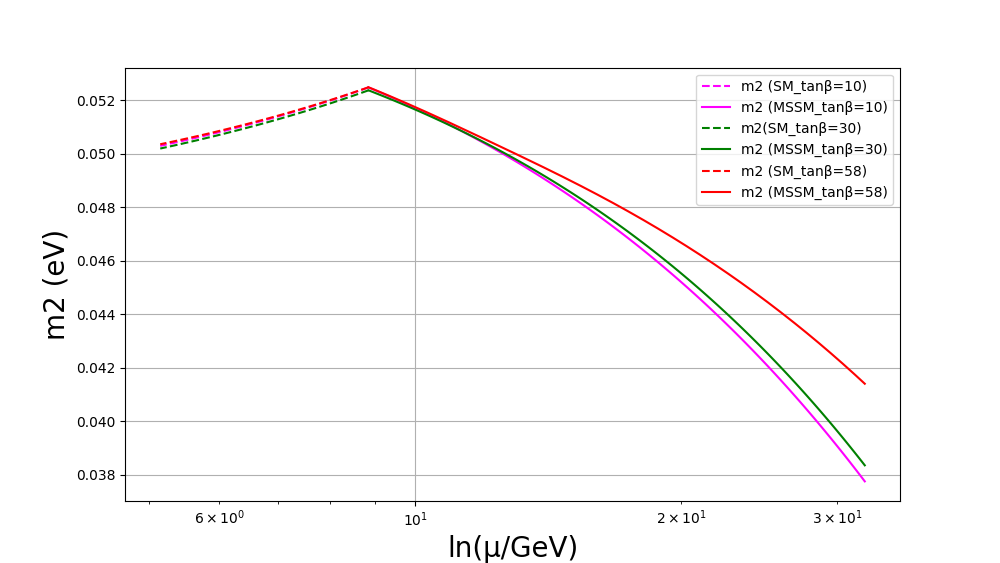}
    \caption{Evolution of $m_2$}
    \label{m2_no1}
  \end{subfigure} \\[2ex]
  \begin{subfigure}[b]{0.45\textwidth}
    \centering
    \includegraphics[height=5cm]{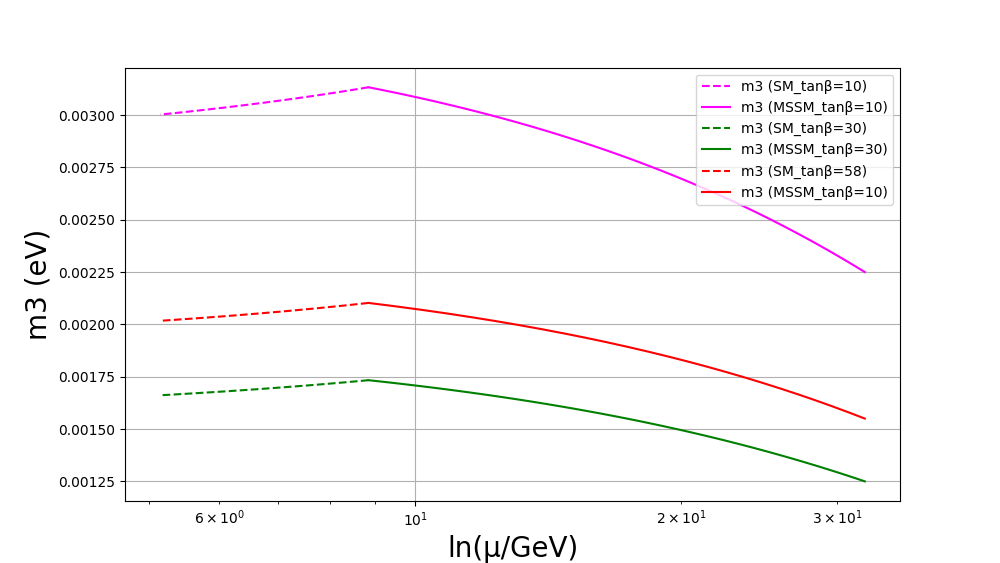}
    \caption{Evolution of $m_3$}
    \label{m3_no1}
  \end{subfigure} &
  \begin{subfigure}[b]{0.45\textwidth}
    \centering
    \includegraphics[height=5cm]{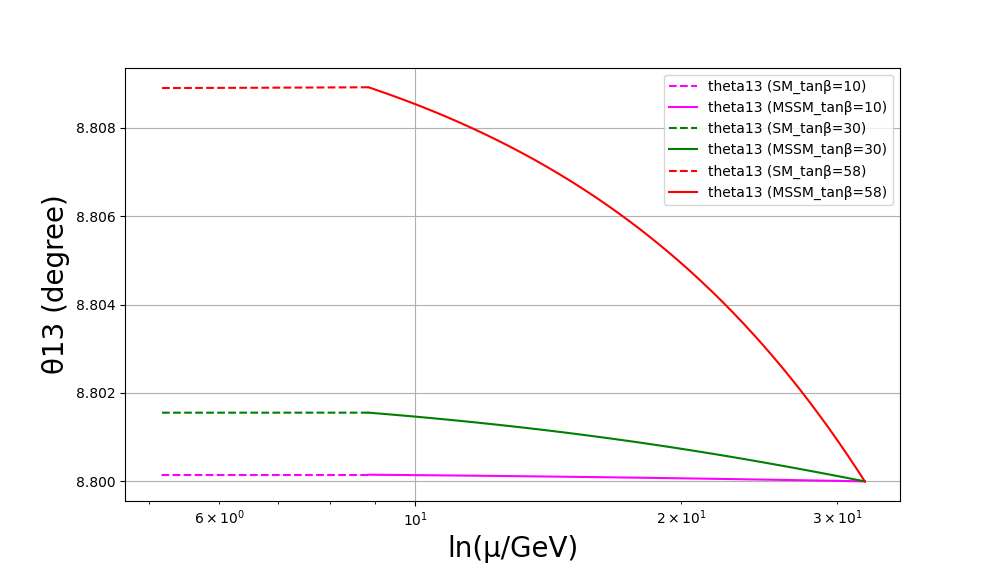}
    \caption{Evolution of $\theta_{13}$}
    \label{theta12_no1}
  \end{subfigure} \\[2ex]
  \begin{subfigure}[b]{0.45\textwidth}
    \centering
    \includegraphics[height=5cm]{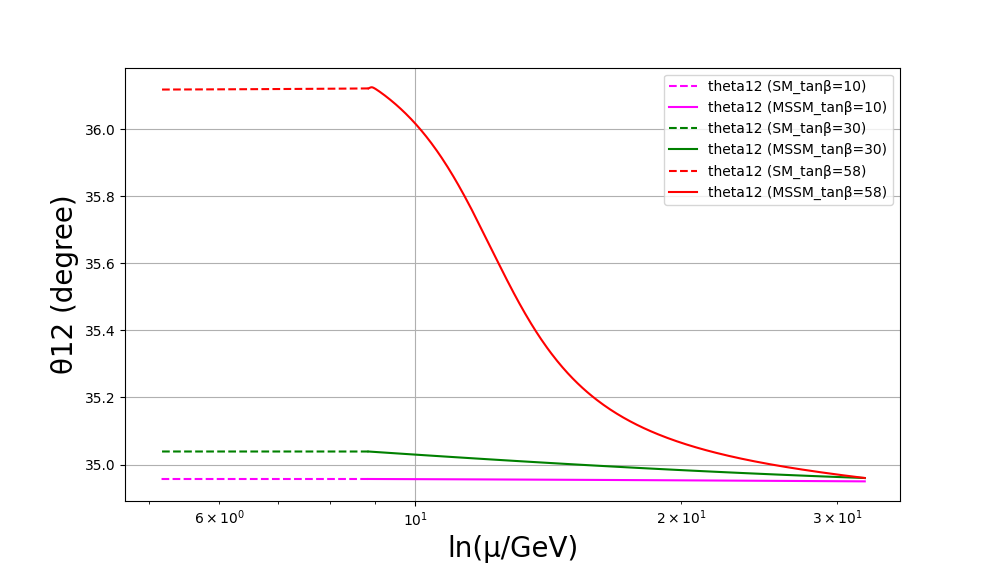}
    \caption{Evolution of $\theta_{12}$}
    \label{theta13_no1}
  \end{subfigure} &
  \begin{subfigure}[b]{0.45\textwidth}
    \centering
    \includegraphics[height=5cm]{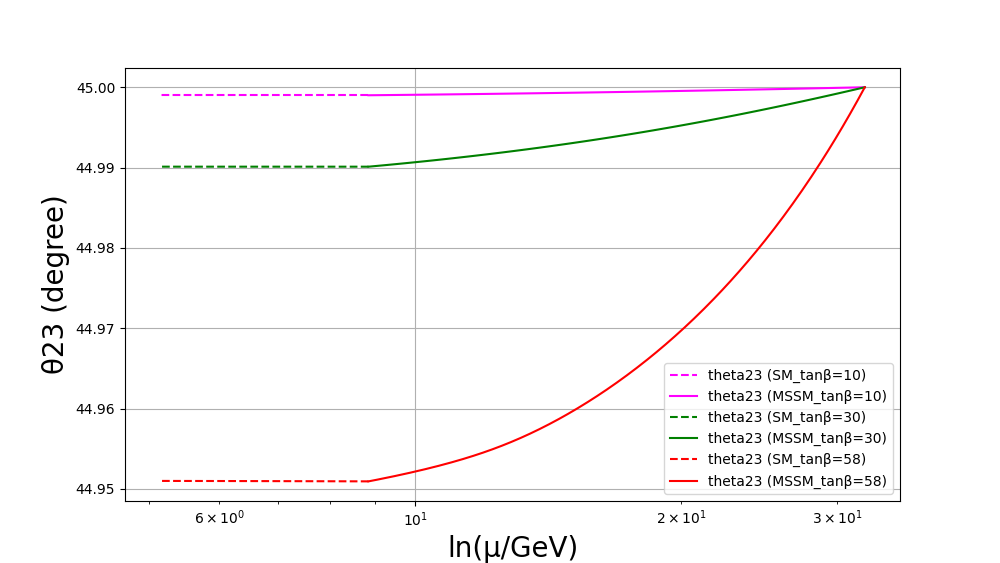}
    \caption{Evolution of $\theta_{23}$}
    \label{theta23_no1}
  \end{subfigure} \\[2ex]
    \end{tabular}
  \caption{Evolution of mass eigenvalues and mixing angles with energy scale for IO and case-II with three different values of $\tan\beta$. Solid and dashed portions of each curve represent the evolution in the MSSM and SM regions respectively. Magenta , green and red colors respectively stands for $\tan\beta=10,\  30\ \text{and}\ 58$.}
  \label{FIO21}
\end{figure}

\begin{figure}[!t]
  \centering
  \begin{tabular}{cc}
  \begin{subfigure}[b]{0.45\textwidth}
    \centering
    \includegraphics[height=5cm]{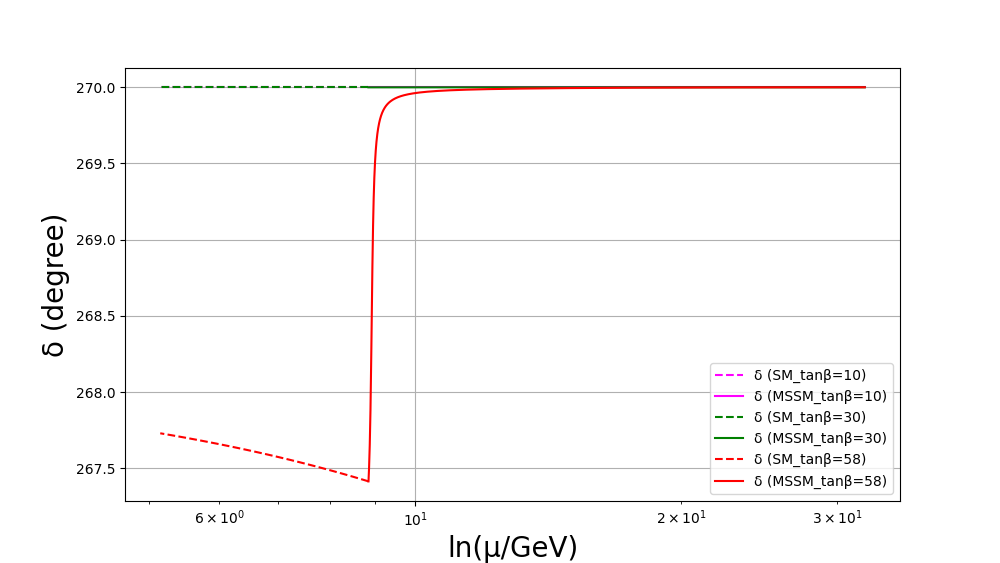}
    \caption{Evolution of $\delta$}
    \label{delta_no1}
  \end{subfigure} &
  \begin{subfigure}[b]{0.45\textwidth}
    \centering
    \includegraphics[height=5cm]{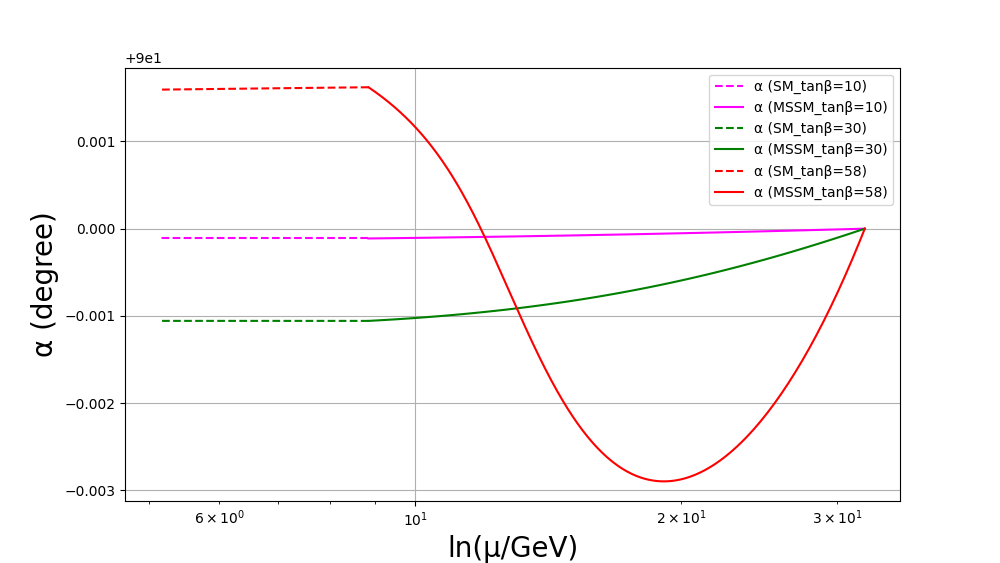}
    \caption{Evolution of $\alpha$}
  \end{subfigure} \\[2ex]
  \begin{subfigure}[b]{0.45\textwidth}
    \centering
    \includegraphics[height=5cm]{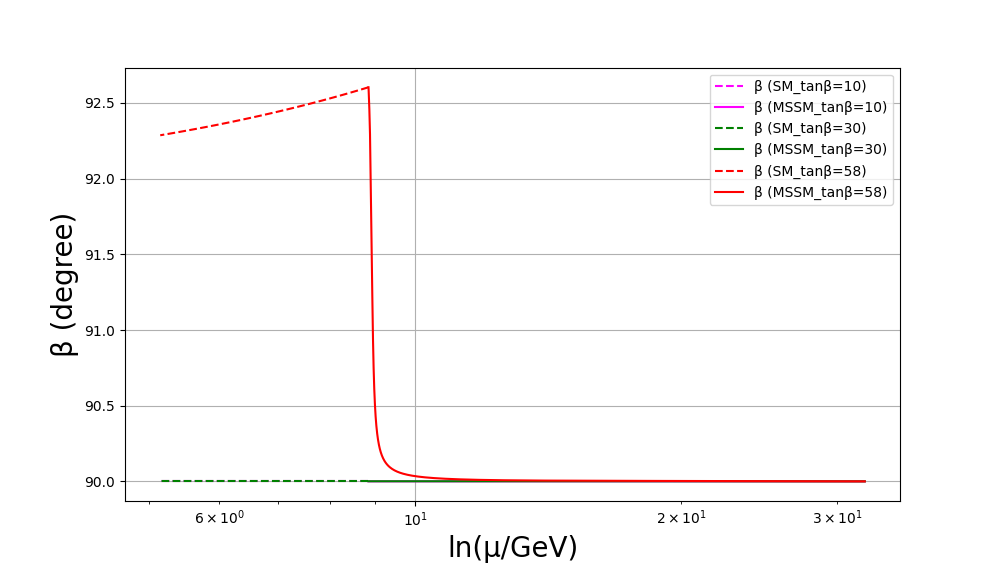}
    \caption{Evolution of $\beta$ }
  \end{subfigure} &
  \begin{subfigure}[b]{0.45\textwidth}
    \centering
    \includegraphics[height=5cm]{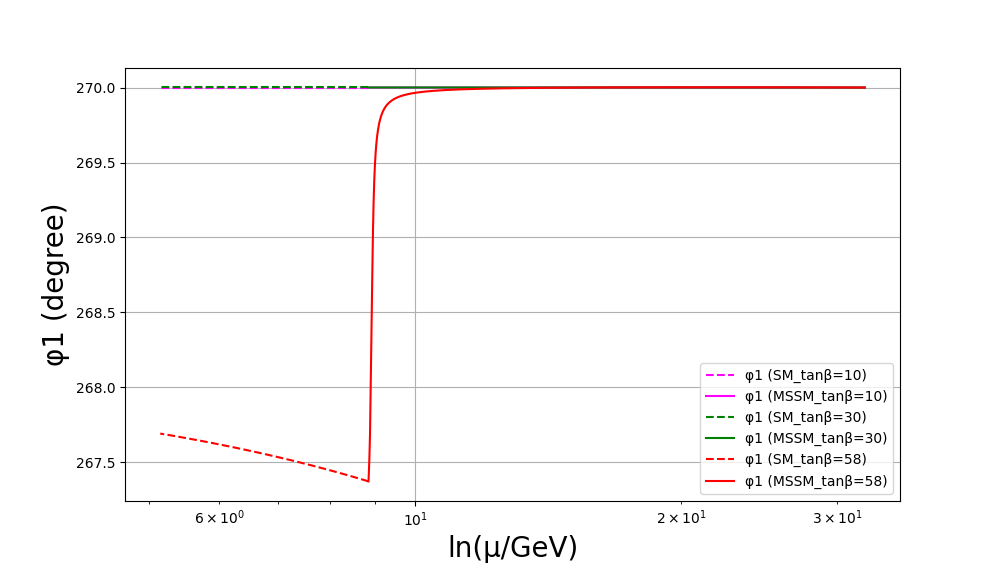}
    \caption{Evolution of $\phi_1$}
  \end{subfigure} \\[2ex]
  \begin{subfigure}[b]{0.45\textwidth}
    \centering
    \includegraphics[height=5cm]{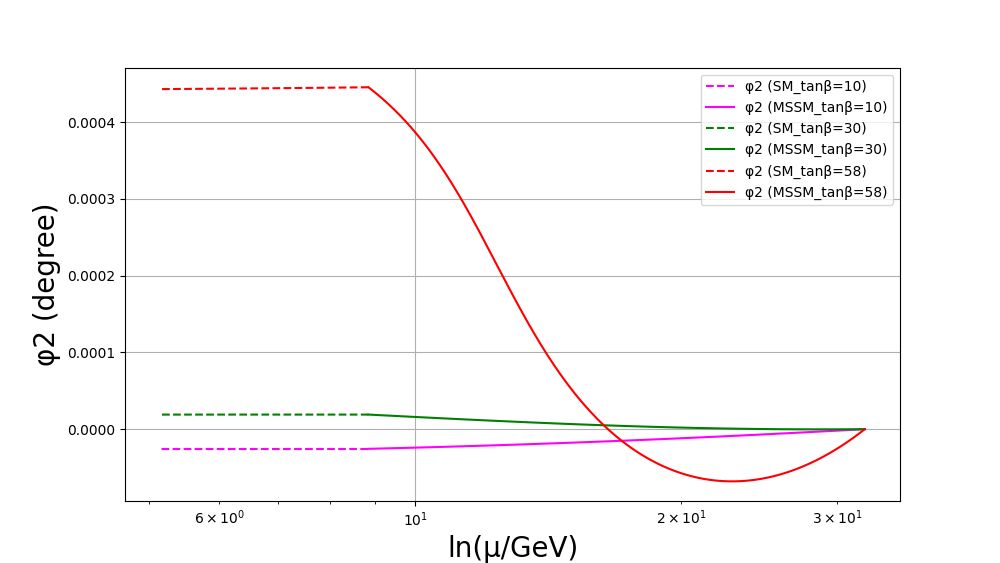}
    \caption{Evolution of $\phi_2$}
  \end{subfigure} & 
  \begin{subfigure}[b]{0.45\textwidth}
    \centering
    \includegraphics[height=5cm]{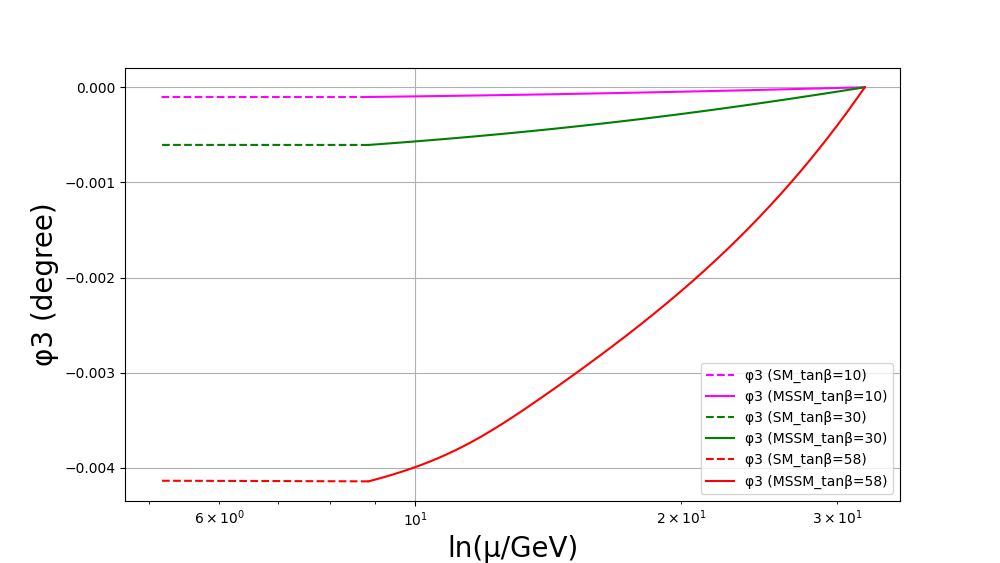}
    \caption{Evolution of $\phi_3$}
    \label{phi3_no1}
  \end{subfigure}  
  \end{tabular}
  \caption{Evolution of CP phases with energy scale for IO and case-II with three different values of $\tan\beta$. Solid and dashed portions of each curve represent the evolution in the MSSM and SM regions respectively. Magenta, green and red colors respectively stands for $\tan\beta=10,\  30\ \text{and}\ 58$.}
  \label{FIO22}
\end{figure}

The results obtained after RG running are presented in Tables (\ref{TNO1}–\ref{TIO2}). Tables \ref{TNO1} and \ref{TNO2} correspond to the normal mass ordering, while Tables \ref{TIO1} and \ref{TIO2} correspond to the inverted mass ordering. The tables list the input values of all parameters at $\Lambda_{FS}$ and their corresponding output values at the top-quark mass scale $m_t$, for three representative values of $\tan\beta$ ($=10,\ 30$ and $58$) but a fixed value of $\Lambda_s\ (=7\ TeV)$. The running behaviors of all parameters are illustrated in Figs. (\ref{FNO11}–\ref{FIO22}). In all figures, the evolution curves shown in magenta, green, and red correspond to $\tan\beta = 10$, $30$, and $58$, respectively. Furthermore, the solid and dashed portions of each curve represent the evolution in the MSSM and SM regions, respectively. The numerical analysis for both cases of NO and IO is done separately and the results are discussed in the following two subsections.
\subsection{Normal Order ($m_1<m_2<m_3$)}
\begin{table}[t]
\begin{center}
\begin{tabular}{c cc cc cc}
\hline
\multirow{2}{*}{Parameter}& 
\multicolumn{2}{c}{$\tan\beta=10$}&
\multicolumn{2}{c}{$\tan\beta=30$}&
\multicolumn{2}{c}{$\tan\beta=58$} \\
\cline{2-7}
 &\makecell{Input at \\ $\Lambda_{FS}$}  & \makecell{Output at \\$\Lambda_{EW}$} &\makecell{Input at \\ $\Lambda_{FS}$}&\makecell{Output at \\ $\Lambda_{EW}$ }&\makecell{Input at \\ $\Lambda_{FS}$}&\makecell{ Output at \\ $\Lambda_{EW}$} \\ \hline
 $m_1\ (eV)$& 0.01109 & 0.014799 & 0.01089 & 0.014473 & 0.01021 & 0.013263 \\
 $m_2\ (eV)$& 0.01295& 0.017258 & 0.01295 & 0.017506 & 0.01295 & 0.015787 \\
 $m_3\ (eV)$& 0.03983 & 0.053158 & 0.03983 & 0.053026 & 0.03993 & 0.015787 \\
$\theta_{13} (/^\circ)$&8.57 & 8.5698 & 8.57 & 8.5688 & 8.57 & 8.5647\\
$\theta_{12} (/^\circ)$& 34 & 33.9971 & 34 & 33.9731 & 34 & 33.8534 \\
$\theta_{23} (/^\circ)$& 45 & 45.0018 & 45 & 45.0176 & 45 & 45.0911 \\
 $\delta (/^\circ)$& 90 & 89.9999 & 90 & 90.0006 & 90 & 90.0227 \\
 $\alpha (/^\circ)$& 270 & 270.0011 & 270 & 270.0103 & 270 & 270.0397 \\
 $\beta (/^\circ)$& 270 & 269.9992 & 270 & 269.9538 & 270 & 269.9629 \\
 $\phi_1 (/^\circ)$& 90 & 89.9993 & 90 & 89.9946 & 90 &89.9955 \\
 $\phi_2 (/^\circ)$& 0 &  -0.000034 & 0 &-0.0003 & 0 & -0.0027 \\
 $\phi_3 (/^\circ)$& 0 &0.000779 & 0 &0.0076 & 0 & 0.0412\\
 $g_1$& 0.625793 & 0.461241 & 0.625793 & 0.461241 & 0.625793 & 0.461241 \\
 $g_2$& 0.684943 & 0.662409 & 0.684943 & 0.662409 & 0.684943 & 0.662409 \\
 $g_3$& 0.722027 & 1.194212 & 0.721910 & 1.193682 & 0.721910 & 1.193682\\
 $y_t$& 0.634446 & 0.997831 & 0.640259 & 0.994130 & 0.700936 & 0.990449 \\
 $y_b$& 0.065629 & 0.002721 & 0.212362 & 0.015791 & 0.601777& 0.091703\\
 $y_{\tau}$& 0.080029 & 0.001726 & 0.259488 & 0.010053 & 0.735524 & 0.058534\\
 $\Delta m^2_{21}(10^{-5}eV^2)$&- & 7.88 & - & 7.93 &- & 7.33 \\
 $\Delta m^2_{32}(10^{-3}eV^2)$&- & 2.52 & - & 2.52 &- & 2.52 \\
 $\sum_i m_i (eV)$&- & 0.08521 & - & 0.08449 &- & 0.08169 \\
 \hline
\end{tabular}
\end{center}
\caption{Input values at $\Lambda_{FS}$ and corresponding low energy values at $m_t$ scale of all the parameters for three different values of $\tan\beta=10, 30$ and $58$ in NO and case-I. }
\label{TNO1}
\end{table} 

For the normal ordering (NO) scenario, the mass eigenvalues at the high-energy scale $\Lambda_{FS}$ are chosen such that $m_1 < m_2 < m_3$. Based on the predictions of the reflection symmetry, there exist two possible values of the Dirac CP phase $\delta$. Accordingly, we consider two distinct cases for our analysis: Case I with $\theta_{23} = \pi/4$ and $\delta = \pi/2$, and Case II with $\theta_{23} = \pi/4$ and $\delta = 3\pi/2$. The numerical evolution is performed separately for each case.

For Case I, the numerical results obtained by solving the complete set of RGEs are summarized in Table \ref{TNO1}. The variations of the mass eigenvalues and mixing angles with respect to the energy scale are illustrated in Figs. 1 (a)–(f), while the evolutions of the CP-violating phases are shown in Figs. 2 (a)–(f). Each plot demonstrates the running behavior of the corresponding parameter for three different values of $\tan\beta$.

As an illustration, let us discuss the numerical analysis and results corresponding to the case of $\tan\beta = 10$. Since the mass eigenvalues at $\Lambda_{\text{FS}}$ are treated as free parameters, they must be carefully chosen to satisfy the low-energy experimental constraints. Even a slight variation in their values at $\Lambda_{\text{FS}}$ can cause significant changes in the mass-squared differences and the total neutrino mass sum. Within the allowed cosmological bound on $\sum m_i$, it is observed that the mass-squared differences initially approach the experimental best-fit values as the input masses increase, but beyond a certain point, they start deviating again. The optimal set of input mass eigenvalues is thus identified as the one yielding mass-squared differences closest to the experimental best-fit results. Any small deviation, either an increase or a decrease from this optimal set, results in noticeable departures from the best-fit values.

For $\tan\beta = 10$, the chosen input mass eigenvalues are $m_1 = 0.01109~\text{eV} < m_2 = 0.01295~\text{eV} < m_3 = 0.03983~\text{eV}$. The corresponding low-energy mass eigenvalues at the $m_t$ scale are obtained as $m_1 = 0.014799~\text{eV}$, $m_2 = 0.017258~\text{eV}$, and $m_3 = 0.053158~\text{eV}$. These values give a total mass sum of $\sum m_i \approx 0.085~\text{eV}$, which lies well within the cosmological upper limit of $0.12~\text{eV}$ \cite{mbound}. The resulting mass-squared differences are $\Delta m_{21}^2 = 7.88 \times 10^{-5}~\text{eV}^2$ and $\Delta m_{32}^2 = 2.52 \times 10^{-3}~\text{eV}^2$, both of which are consistent with the global best-fit values (see Table \ref{GA}). From Figs. 1(a)–1(c), it is evident that the mass eigenvalues gradually increase as the energy scale runs from high to low values.

Similar to the mass eigenvalues, the mixing angles $\theta_{13}$ and $\theta_{12}$ are also treated as free parameters at the high-energy scale. Their input values are taken as $8.57^\circ$ and $34^\circ$, respectively, so that their evolved values at low energy closely reproduce the experimental best-fit values. At the $m_t$ scale, we obtain $\theta_{12} \approx 33.99^\circ$ and $\theta_{13} \approx 8.56^\circ$, which are in good agreement with the global best-fit results (Table \ref{GA}). The value of $\theta_{12}$ is also well within the predicted range by the recent JUNO experiment \cite{JUNO}. Due to the effects of RG running, both $\theta_{12}$ and $\theta_{13}$ exhibit a mild decrease as the energy scale decreases (Figs. 1(d) and 1(e)). The overall running behavior of the three mixing angles is qualitatively similar; however, while $\theta_{12}$ and $\theta_{13}$ decrease with decreasing energy, $\theta_{23}$ shows a slight increase. The input value of $\theta_{23}$ is taken as $45^\circ$, consistent with the $\mu$–$\tau$ reflection symmetry condition. This angle remains close to its maximal value under RG evolution, with a low-energy value of $\theta_{23} = 45.0180^\circ$. It is worth noting that the current best-fit value of $\theta_{23}$ in the normal ordering scenario is $48.5^\circ$ (without SK data) or $43.3^\circ$ (with SK data), as shown in Table \ref{GA}. Therefore, the low-energy prediction of $\theta_{23}$ obtained through RG running is consistent with the global analysis without SK data, as it slightly exceeds $45^\circ$, indicating a preference for the second octant.

For Case I, the high-energy value of the Dirac CP phase $\delta$ at $\Lambda_{FS}$ is set to $90^\circ$, in accordance with the $\mu$–$\tau$ reflection symmetry. The remaining CP-violating phases are assigned their input values as specified in Eq. (\ref{ch1}). The deviations of all CP phases from their initial (maximal) values during RG evolution are found to be extremely small. For instance, the low-energy value of $\delta$ is obtained as $89.99^\circ$, indicating negligible running effects. The low-energy values of the other CP phases, relative to their respective high-energy inputs, are listed in Table \ref{TNO1}, while their RG evolutions are illustrated in Figs. 2(a)–2(f).

To investigate the impact of RG running on neutrino parameters for higher $\tan\beta$, we extend the analysis to $\tan\beta = 30$ and $58$ within the normal ordering (NO) framework for Case I. The same input values of $m_2$, $\theta_{13}$, and $\theta_{12}$ as those used in the $\tan\beta = 10$ case are retained, while the input values of $m_1$ and $m_3$ are adjusted as listed in Table \ref{TNO1}. With increasing $\tan\beta$, noticeable variations appear in the low-energy predictions of each parameter, as can be seen from Table \ref{TNO1}. Nevertheless, the resulting mass-squared differences and the total mass sum $\sum m_i$ remain consistent with experimental constraints. The degree of separation among the evolution curves in each plot represents the influence of the variation of $\tan\beta$ on the corresponding parameter. From Figs. 1(a)–1(f), it is evident that the running behavior of the mass eigenvalues remains nearly identical across $\tan\beta = 10$, $30$, and $58$, showing only minimal deviations. In contrast, the mixing angles exhibit more pronounced variations, particularly for the case of $\tan\beta = 58$, where the RG running effects become stronger. Similarly, in the case of the CP-violating phases (Figs. 2(a)–2(f)), it is observed that the evolution curves shown in red, corresponding to $\tan\beta = 58$, follow a noticeably distinct trajectory compared to those for $\tan\beta = 10$ and $30$. This indicates that the RG running effects become significantly more pronounced at higher $\tan\beta$, making the behavior of the CP phases for $\tan\beta = 58$ clearly distinguishable from the lower $\tan\beta$ cases.

\begin{table}[t]
\begin{center}
\begin{tabular}{c cc cc cc}
\hline
\multirow{2}{*}{Parameter}& 
\multicolumn{2}{c}{$\tan\beta=10$}&
\multicolumn{2}{c}{$\tan\beta=30$}&
\multicolumn{2}{c}{$\tan\beta=58$} \\
\cline{2-7}
 &\makecell{Input at \\ $\Lambda_{FS}$}  & \makecell{Output at \\$\Lambda_{EW}$} &\makecell{Input at \\ $\Lambda_{FS}$}&\makecell{Output at \\ $\Lambda_{EW}$ }&\makecell{Input at \\ $\Lambda_{FS}$}&\makecell{ Output at \\ $\Lambda_{EW}$} \\ \hline
 $m_1\ (eV)$& 0.01115 & 0.014862 &  0.01135 & 0.014915 & 0.01235 & 0.014829 \\
 $m_2\ (eV)$& 0.01295& 0.017277 & 0.01295 & 0.017187 & 0.01295 & 0.017105 \\
 $m_3\ (eV)$& 0.03963 & 0.052891 &  0.03973& 0.052890 & 0.03993 & 0.052567 \\
$\theta_{13} (/^\circ)$&8.57 & 8.5698 & 8.57 & 8.5688 & 8.57 & 8.5647\\
$\theta_{12} (/^\circ)$& 34 & 33.9929 & 34 & 33.9268 & 34 & 33.5551 \\
$\theta_{23} (/^\circ)$& 45 & 45.0006 & 45 & 45.0064 & 45 & 45.0288 \\
$\delta  (/^\circ)$& 270 & 269.9999 & 270 & 269.9984 & 270 & 269.9735 \\
 $\alpha (/^\circ)$& 90 & 89.9988 & 90 & 89.9899 & 90 & 89.9648 \\
 $\beta (/^\circ)$& 90 & 90.00002 & 90 & 90.0005 & 90 & 90.0135 \\
 $\phi_1 (/^\circ)$& 270 & 270.0006 & 270 & 270.0045 & 270 &269.9978 \\
 $\phi_2 (/^\circ)$& 0 & 0.00024 & 0 &0.0021 & 0 & 0.0049 \\
 $\phi_3 (/^\circ)$& 0 &0.00045  & 0 &0.0045 & 0 & 0.0242\\
 $g_1$& 0.625793 & 0.461241 & 0.625793 & 0.461241 & 0.625793 & 0.461241 \\
 $g_2$& 0.684943 & 0.662409 & 0.684943 & 0.662409 & 0.684943 & 0.662409 \\
 $g_3$& 0.722027 & 1.194212 & 0.721910 & 1.193682 & 0.721910 & 1.193682\\
 $y_t$& 0.634446 & 0.997831 & 0.640259 & 0.994130 & 0.700936 & 0.990449 \\
 $y_b$& 0.065629 & 0.002721 & 0.212362 & 0.015791 & 0.601777& 0.091703\\
 $y_{\tau}$& 0.080029 & 0.001726 & 0.259488 & 0.010053 & 0.735524 & 0.058534\\
 $\Delta m^2_{21}(10^{-5}eV^2)$&- & 7.76 & - & 7.29 &- & 7.26 \\
 $\Delta m^2_{32}(10^{-3}eV^2)$&- & 2.49 & - & 2.50 &- & 2.47 \\
 $\sum_i m_i (eV)$&- & 0.08502 & - & 0.08499 &- & 0.08450 \\
 \hline
\end{tabular}
\end{center}
\caption{Input values at $\Lambda_{FS}$ and corresponding low energy values at $m_t$ scale of all the parameters for three different values of $\tan\beta=10, 30$ and $58$ in NO and case-II. }
\label{TNO2}
\end{table} 

For Case II, the high-energy input value of the Dirac CP-violating phase $\delta$ at $\Lambda_{\text{FS}}$ is set to $270^\circ$, accompanied by $\theta_{23} = 45^\circ$. The remaining CP-violating phases are assigned their input values as given in Eq.~(\ref{ch2}), consistent with the $\mu$–$\tau$ reflection symmetry. The input values of the free parameters are again chosen such that the resulting low-energy predictions satisfy the experimental constraints. As in Case I, the numerical analysis is performed for three different values of $\tan\beta (= 10$, $30$, and $58$) and the obtained results are presented in Table \ref{TNO2}. The evolution of the mass eigenvalues and mixing angles with the energy scale is illustrated in Figs. 3(a)–3(f), while the corresponding RG running behaviors of the CP-violating phases are shown in Figs. 4(a)–4(f). Each plot depicts the variation of the respective parameter for the three chosen values of $\tan\beta$.

The results obtained for Case II exhibit qualitative similarities with those of Case I. The chosen high-energy input values of the mass eigenvalues and the two mixing angles constitute an optimal set that yields low-energy predictions most consistent with experimental observations. Even a slight deviation from this optimal set leads to noticeable discrepancies in the predicted mass-squared differences relative to their global best-fit values. The predicted low-energy sum of neutrino masses, $\sum m_i \approx 0.089~\text{eV}$, remains well within the cosmological upper bound. The RG running behavior of the mass eigenvalues, shown in Figs. 3(a)–3(c) closely follow the trend observed in Case I — all three mass eigenvalues increase gradually as the energy scale decreases.

Similar to Case I, the RG running of the mixing angles and CP-violating phases in Case II is also found to be relatively mild, leading to only small deviations of these parameters from their initial values. Nevertheless, the low-energy predictions remain in good agreement with the global best-fit results (Table \ref{GA}). The value of $\theta_{12}$ is found to be within the range as predicted by recent JUNO experiment for NO case II also \cite{JUNO}. The behavior of $\theta_{23}$ under RG evolution is the same as observed in Case I—it slightly increases with decreasing energy scale and remains in the second octant, with a low-energy predicted value of $\theta_{23} = 45.03^\circ$ for all three choices of $\tan\beta$. All CP-violating phases, including the Dirac phase $\delta$, exhibit only minor deviations due to RG running. The corresponding low-energy values of the CP phases at the $m_t$ scale can be found in Table \ref{TNO2}. 

The energy evolution curves of the parameters generated for the three choices of $tan\beta$ in case II exhibit similar characteristics as observed in case I. For the mass eigenvalues (Figs 3(a)-(c)), all three curves corresponding to $tan\beta=10,\ 30$ and $58$ are relatively closer, implying that the impact of variation of $tan\beta$ is less significant. On the other hand, for mixing angles and CP phases running effects corresponding to $tan\beta=58$ relatively differ from those corresponding to $tan\beta=10$ and $30$ (evolution curve in red is distinct from the other two in Figs. 3(d)-(f) and 4(a)-(f)).

\begin{table}[t]
\begin{center}
\begin{tabular}{c cc cc cc}
\hline
\multirow{2}{*}{Parameter}& 
\multicolumn{2}{c}{$\tan\beta=10$}&
\multicolumn{2}{c}{$\tan\beta=30$}&
\multicolumn{2}{c}{$\tan\beta=58$} \\
\cline{2-7}
 &\makecell{Input at \\ $\Lambda_{FS}$}  & \makecell{Output at \\$\Lambda_{EW}$} &\makecell{Input at \\ $\Lambda_{FS}$}&\makecell{Output at \\ $\Lambda_{EW}$ }&\makecell{Input at \\ $\Lambda_{FS}$}&\makecell{ Output at \\ $\Lambda_{EW}$} \\ \hline
 $m_1\ (eV)$& 0.03717 & 0.049587 & 0.03793 & 0.049824 & 0.03791 & 0.049277 \\
 $m_2\ (eV)$& 0.03775& 0.050329 & 0.03805 & 0.050604 & 0.04110 & 0.050062 \\
 $m_3\ (eV)$&0.00225 & 0.003002 & 0.00125 & 0.001661 & 0.00115 & 0.001517 \\
$\theta_{13} (/^\circ)$&8.8 & 8.8001 & 8.46 & 8.8015 & 8.57 & 8.4690\\
$\theta_{12} (/^\circ)$& 34.95 & 35.0019 & 34.37 & 35.8545 & 34 & 33.9849 \\
$\theta_{23} (/^\circ)$& 45 & 44.9989 & 45 & 44.9931 & 45 & 44.9416 \\
 $\delta (/^\circ)$& 90 & 90.00004 & 90 & 90.0032 & 90 & 93.1178 \\
 $\alpha (/^\circ)$& 270 & 270.0009 & 270 & 270.0025 & 270 & 266.8990 \\
 $\beta (/^\circ)$& 270 & 269.9998 & 270 & 269.9948 & 270 & 269.9851 \\
 $\phi_1 (/^\circ)$& 90 & 89.9994 & 90 & 90.0099 & 90 &93.1175\\
 $\phi_2 (/^\circ)$& 0 &  -0.0001 & 0 &-0.0013 & 0 & -0.0007 \\
 $\phi_3 (/^\circ)$& 0 &0.0002 & 0 &0.0021 & 0 & 0.0075\\
 $g_1$& 0.625793 & 0.461241 & 0.625793 & 0.461241 & 0.625793 & 0.461241 \\
 $g_2$& 0.684943 & 0.662409 & 0.684943 & 0.662409 & 0.684943 & 0.662409 \\
 $g_3$& 0.722027 & 1.194212 & 0.721910 & 1.193682 & 0.721910 & 1.193682\\
 $y_t$& 0.634446 & 0.997831 & 0.640259 & 0.994130 & 0.700936 & 0.990449 \\
 $y_b$& 0.065629 & 0.002721 & 0.212362 & 0.015791 & 0.601777& 0.091703\\
 $y_{\tau}$& 0.080029 & 0.001726 & 0.259488 & 0.010053 & 0.735524 & 0.058534\\
 $\Delta m^2_{21}(10^{-5}eV^2)$&- & 7.41 & - & 7.83 &- & 7.79 \\
 $\Delta m^2_{32}(10^{-3}eV^2)$&- & 2.52 & - & 2.55 &- & 2.50 \\
 $\sum_i m_i (eV)$&- & 0.10291 & - & 0.10208 &- & 0.10085 \\
 \hline
\end{tabular}
\end{center}
\caption{Input values at $\Lambda_{FS}$ and corresponding low energy values at $m_t$ scale of all the parameters for three different values of $\tan\beta=10, 30$ and $58$ in IO and case-I. }
\label{TIO1}
\end{table} 
\subsection{Inverted Order ($m_3<m_1<m_2$)}
We have carried out a similar analysis for the inverted ordering scenario as was performed for the normal ordering case. In this framework, we searched for the high-energy input values of the free parameters that can reproduce low-energy predictions consistent with the current observational data. Through a detailed investigation, an optimal set of input parameters has been identified, leading to low-energy predictions that lie within the $3\sigma$ allowed ranges of the global-fit data. We first present the numerical results for Case I, which corresponds to the maximal predictions $\theta_{23}=45^\circ$ and $\delta=90^\circ$ as implied by the $\mu$–$\tau$ reflection symmetry.

For the inverted ordering scenario with $\tan\beta = 10$, we have chosen the high-energy input values of the mass eigenvalues at the flavor symmetry scale $\Lambda_{FS}$ as $m_3 = 0.00225~\text{eV} < m_1 = 0.03717~\text{eV} < m_2 = 0.03775~\text{eV}.$
After renormalization group (RG) evolution down to the top-quark mass scale ($m_t$), the corresponding low-energy mass eigenvalues are obtained as 
$m_3 = 0.003002~\text{eV} < m_1 = 0.049587~\text{eV} < m_2 = 0.050329~\text{eV}.$ These low-energy values yield a total neutrino mass sum of 
$\sum m_i \approx 0.1029~\text{eV},$ which satisfies the cosmological upper bound of $0.12~\text{eV}$. The corresponding mass-squared differences are found to be 
$\Delta m^2_{21} = 7.41\times 10^{-5}~\text{eV}^2$ and $\Delta m^2_{32} = -2.52\times 10^{-3}~\text{eV}^2,$ both of which lie well within the $3\sigma$ allowed ranges of the global analysis data (Table~\ref{GA}).

The high-energy input values of the mixing angles $\theta_{12}$ and $\theta_{13}$ at  $\Lambda_{FS}$ are taken as $34.95^\circ$ and $8.8^\circ$, respectively, for the case of $\tan\beta = 10$. Due to renormalization group (RG) running, they evolve to $\theta_{12} = 35.0019^\circ$ and $\theta_{13} = 8.8001^\circ$ at the top-quark mass scale ($m_t$). As observed in the NO scenario, the deviation of $\theta_{23}$ from its maximal value in the present case of inverted ordering (IO) is also very small. Its low-energy value is found to be approximately $44.99^\circ$, exhibiting a tendency to lie in the first octant. This behavior contrasts with the NO scenario, where $\theta_{23}$ tends to lie in the second octant due to RG running. Nonetheless, the low-energy values of all three mixing angles at the $m_t$ scale remain consistent with the $3\sigma$ allowed ranges of the global analysis data (Table~\ref{GA}). Similar to the NO scenario, the RG running of the CP-violating phases in the IO case is also found to be relatively weak, leading to only small deviations from their high-energy values.

\begin{table}[t]
\begin{center}
\begin{tabular}{c cc cc cc}
\hline
\multirow{2}{*}{Parameter}& 
\multicolumn{2}{c}{$\tan\beta=10$}&
\multicolumn{2}{c}{$\tan\beta=30$}&
\multicolumn{2}{c}{$\tan\beta=58$} \\
\cline{2-7}
 &\makecell{Input at \\ $\Lambda_{FS}$}  & \makecell{Output at \\$\Lambda_{EW}$} &\makecell{Input at \\ $\Lambda_{FS}$}&\makecell{Output at \\ $\Lambda_{EW}$ }&\makecell{Input at \\ $\Lambda_{FS}$}&\makecell{ Output at \\ $\Lambda_{EW}$} \\ \hline
 $m_1\ (eV)$& 0.03711 & 0.049542 & 0.03703 & 0.0494 & 0.03761 & 0.049612 \\
 $m_2\ (eV)$& 0.03775& 0.050293 & 0.03835 & 0.050193 & 0.04140 & 0.050353 \\
 $m_3\ (eV)$& 0.00225 & 0.003002 & 0.00125 & 0.001661 & 0.00155 & 0.002017 \\
$\theta_{13} (/^\circ)$&8.8 & 8.8001 & 8.8 & 8.8015 & 8.8 & 8.8088\\
$\theta_{12} (/^\circ)$& 34.95 & 34.9570 & 34.96 & 35.0390 & 34.96 & 36.1178 \\
$\theta_{23} (/^\circ)$& 45 & 44.9990 & 45 & 44.9901 & 45 & 44.9509 \\
$\delta  (/^\circ)$& 270 & 269.9999 & 270 & 269.9996 & 270 & 267.7294 \\
 $\alpha (/^\circ)$& 90 & 89.9998 & 90 &  89.9989 & 90 & 90.0015 \\
 $\beta (/^\circ)$& 90 & 90.0001 & 90 & 90.0007 & 90 & 92.2866 \\
 $\phi_1 (/^\circ)$& 270 & 270.0001 & 270 & 270.0006 & 270 &267.6918 \\
 $\phi_2 (/^\circ)$& 0 & -0.00002 & 0 &0.00001 & 0 & 0.0004 \\
 $\phi_3 (/^\circ)$& 0 & -0.0001  & 0 &-0.0006 & 0 & -0.0041\\
 $g_1$& 0.625793 & 0.461241 & 0.625793 & 0.461241 & 0.625793 & 0.461241 \\
 $g_2$& 0.684943 & 0.662409 & 0.684943 & 0.662409 & 0.684943 & 0.662409 \\
 $g_3$& 0.722027 & 1.194212 & 0.721910 & 1.193682 & 0.721910 & 1.193682\\
 $y_t$& 0.634446 & 0.997831 & 0.640259 & 0.994130 & 0.700936 & 0.990449 \\
 $y_b$& 0.065629 & 0.002721 & 0.212362 & 0.015791 & 0.601777& 0.091703\\
 $y_{\tau}$& 0.080029 & 0.001726 & 0.259488 & 0.010053 & 0.735524 & 0.058534\\
 $\Delta m^2_{21}(10^{-5}eV^2)$&- & 7.50 & - & 7.61 &- & 7.41 \\
 $\Delta m^2_{32}(10^{-3}eV^2)$&- & 2.52 & - & 2.51 &- & 2.53 \\
 $\sum_i m_i (eV)$&- & 0.10283 & - & 0.10128 &- &  0.10198 \\
 \hline
\end{tabular}
\end{center}
\caption{Input values at $\Lambda_{FS}$ and corresponding low energy values at $m_t$ scale of all the parameters for three different values of $\tan\beta=10, 30$ and $58$ in IO and case-II. }
\label{TIO2}
\end{table} 

To investigate the impact of varying $\tan\beta$ on the running parameters, we have studied the low-energy output values of all the parameters for the cases $\tan\beta = 30$ and $58$, using the same high-energy input values as given in Table~\ref{TIO1}. A significant impact on the low-energy predictions is observed with the increase in the value of $\tan\beta$. All the chosen input parameters are found to reproduce the low-energy constraints after RG evolution, as summarised in Table~\ref{TIO1}. 

The evolution of the mass eigenvalues and mixing angles with energy scale is illustrated in Figs.~5(a)-(f), while the corresponding evolution of the CP-violating phases is shown in Figs.~6(a)--(f). From Figs.~5(a)-(b), it is evident that the three curves in each figure almost overlap, indicating that the variation in the value of $tan\beta$ does not induce any significant change in the running behavior of the mass eigenvalues. However, Figs.~5(d)--(f) reveal that the dependence of RG running of the mixing angles on $\tan\beta$ exhibits a peculiar behavior in the IO Case~I scenario. A similar pattern of distinct splitting and variation in the low-energy output values can also be observed for the CP phases $\beta$, $\phi_2$, and $\phi_3$ in Figs.~6(c), (e), and (f), respectively. On the other hand, the variations in the low-energy output values of $\delta$, $\alpha$, and $\phi_1$ are found to be more pronounced, as shown in Figs.~6(a), (b), and (d).

For Case~II in the inverted ordering (IO) scenario, the high-energy input values of the Dirac CP phase and the atmospheric mixing angle are taken as $\delta = 270^\circ$ and $\theta_{23} = 45^\circ$, respectively, while the other CP phases are given by Eq.~(\ref{ch2}). In this case as well, it is highly challenging to select an appropriate set of input values for the free parameters that can lead to phenomenologically consistent results under RG running. Nevertheless, an optimal set of input parameters has been identified, yielding low-energy predictions that are in good agreement with experimental observations and exhibiting smooth evolution of all parameters without distortions. The chosen high-energy input mass eigenvalues are $m_3 = 0.00225~\text{eV} < m_1 = 0.03711~\text{eV} < m_2 = 0.03775~\text{eV},$ for $\tan\beta = 10$. In addition, the input mixing angles are taken as $\theta_{12} = 34.95^\circ$ and $\theta_{13} = 8.8^\circ$. It is worth noting that even a small variation in these input values leads to significant inconsistencies in the desired low-energy results. With this specific set of inputs, the low-energy neutrino masses yield a total mass sum of $\sum m_i \approx 0.10283~\text{eV},$ which satisfies the cosmological upper bound. Furthermore, the corresponding low-energy mass-squared differences are obtained as $\Delta m^2_{21} = 7.50\times 10^{-5}~\text{eV}^2$ and $\Delta m^2_{32} = -2.52\times 10^{-3}~\text{eV}^2$ for $tan\beta = 10$, both of which lie within the $3\sigma$ ranges of the global analysis data (Table~\ref{GA}). The predicted low-energy values of $\theta_{12} \approx 34.957^\circ$ and $\theta_{13} \approx 8.8001^\circ$ are also consistent with the $3\sigma$ ranges. The atmospheric angle $\theta_{23}$ is found to slightly decrease with the energy scale, settling in the first octant with a predicted value of $\theta_{23} \approx 44.999^\circ$. This running behavior of $\theta_{23}$ closely resembles that observed in Case~I of the IO scenario.

Since the results from the T2K and NO$\nu$A experiments, as well as the global analysis of oscillation data, indicate a maximal Dirac CP phase around $270^\circ$ in the IO scenario, it is important to examine the running behavior of $\delta$ in the present case. According to the latest global analysis, the best-fit value of $\delta$ is $285^\circ$ (without SK data) or $274^\circ$ (with SK data). We find that under RG evolution, $\delta$ exhibits a slight decrease with the energy scale, attaining a low-energy value of $\delta \approx 269.99^\circ$, which remains consistent with the $3\sigma$ allowed range. Similar to $\delta$, the other CP phases also experience only small perturbations during RG running.

Similar to the previous analysis, we examine the impact of varying $\tan\beta$ on the low-energy output values of the parameters for the IO Case II scenario, considering $\tan\beta = 30$ and $58$. As in Case~I, taking $\tan\beta = 58$ leads to noticeable variations in the mixing angles and CP-violating phases. In contrast, for $\tan\beta = 10$ and $30$, the corresponding plots nearly overlap, indicating that the variation of $\tan\beta$ within this range produces no significant effect on the evolution of the parameters. The low-energy predictions of all the parameters are summarised in Table~\ref{TIO2}, and they are found to be consistent with the $3\sigma$ allowed ranges of the global analysis data. Regarding the effects of varying the SUSY-breaking scale on the running of parameters in this case, the behavior remains similar to that observed in Case~I. In particular, the mass eigenvalues and the mixing angles $\theta_{13}$ and $\theta_{23}$ show comparable patterns of evolution under RG running.

\section{Summary and discussion}
The experimentally observed near-maximal value of the atmospheric mixing angle $\theta_{23}$ provides strong motivation to investigate the framework of $\mu$–$\tau$ reflection symmetry. In addition, recent indications of a nearly maximal Dirac CP-violating phase $\delta$ from the T2K and NO$\nu$A experiments further strengthen this motivation. While experiments suggest the values of $\theta_{23}$ and $\delta$ close to maximality, the theoretical prediction arising from exact $\mu$–$\tau$ reflection symmetry yields precisely maximal values. It says the importance of exploring possible mechanisms responsible for breaking the symmetry, and renormalization group evolution (RGE) is a compelling one. Guided by this intuition, we have examined the breaking of $\mu$–$\tau$ reflection symmetry through the RG evolution of neutrino parameters.

Our working premise is that the symmetry holds exactly at a very high energy scale, denoted as $\Lambda_{FS}$. As the energy scale is lowered, the neutrino parameters evolve according to their respective RGEs, which form a set of coupled differential equations that must be solved simultaneously to track their energy dependence. We adopt the MSSM as the effective theory above the supersymmetry breaking scale and the SM below it. In this context, two parameters play a crucial role: the SUSY-breaking scale and $\tan\beta$, both of which remain experimentally unconstrained. Their values significantly affect the RG running behaviour.

In our earlier work \cite{CBorah}, we investigated the impact of varying the supersymmetry breaking scale while keeping $\tan\beta$ fixed. In the present study, we extend this analysis by exploring a range of possible $\tan\beta$ values for a fixed SUSY-breaking scale, thereby providing a more comprehensive understanding of the RG-induced breaking of $\mu$–$\tau$ reflection symmetry.

In this work, we employ the RGEs for the neutrino parameters derived in our earlier study. The RG running is performed from the high-energy flavor symmetry scale $\Lambda_{FS}$ down to the electroweak scale $\Lambda_{EW}$. Since $\mu$–$\tau$ reflection symmetry is assumed to be exactly preserved at $\Lambda_{FS}$, it imposes the conditions that $\theta_{23}$ and the Majorana phases take their maximal values, as discussed in Section~2. In contrast, the three mass eigenvalues along with the mixing angles $\theta_{12}$ and $\theta_{13}$ remain unconstrained at this scale. Their values are chosen appropriately at $\Lambda_{FS}$ so that, after RG evolution, they reproduce the experimentally allowed ranges at low energies.

The analysis is carried out for both normal and inverted mass orderings. For each ordering, we consider three representative values of $\tan\beta$, namely $10$, $30$, and $58$, to investigate the impact of $\tan\beta$ on the RG evolution, while fixing the supersymmetry breaking scale at $\Lambda_s = 7 ,\text{TeV}$. In all cases, the running of the mass eigenvalues is found to exhibit negligible sensitivity to variations in $\tan\beta$. However, the mixing angles and CP-violating phases display significant changes, particularly for larger values of $\tan\beta$. Thus, our analysis reinforces the well-established observation that higher values of $\tan\beta$ lead to more pronounced RG running effects. Furthermore, the relatively small deviations in the neutrino parameters after RG running indicate that $\mu$–$\tau$ reflection symmetry remains largely stable under RG evolution. The selected values of the five free parameters at the high-energy scale are shown to reproduce the low-energy experimental constraints successfully, remaining well within the $3\sigma$ ranges.

\end{document}